\documentclass[12pt]{article}
\usepackage{amsmath}
\usepackage{amssymb}
\usepackage{epsfig}
\usepackage{amsthm}

\numberwithin{equation}{section}
\numberwithin{figure}{section}

\def\eq#1{(\ref{eq:#1})}
\def\lineup{\!\!\!\!\!\!\!\!&&}

\newcommand{\?}{\stackrel{?}{=}}
\def\d{\partial}

\def\Box#1{\boxed{\phantom{\Bigg)}#1\ \ \ }}

\def\eps{\epsilon}
\def\beps{\bar{\epsilon}}

\def\A{\mathcal{A}}
\def\D{\mathcal{D}}
\def\Img{\mathrm{Im}\,}
\def\ker{\mathrm{ker}\,}

\def\integrability{consistency }

\newtheorem{fact2}{Fact}

\begin{document}

\begin{titlepage}

\begin{center}

\vskip 1.0cm {\large \bf{Connecting Solutions in Open String Field Theory}}
\vskip .2cm
{\large \bf{with Singular Gauge Transformations }}\\
\vskip 1.5cm
{\large Theodore Erler$^{(a)}$\footnote{Email: tchovi@gmail.com}}
\vskip .3cm

{\large Carlo Maccaferri$^{(a,b)}$\footnote{Email: maccafer@gmail.com}}

\vskip 1.0cm

$^{(a)}${\it {Institute of Physics of the ASCR, v.v.i.} \\
{Na Slovance 2, 182 21 Prague 8, Czech Republic}}\\
\vskip .4cm

$^{(b)}${\it INFN, Sezione di Torino\\
Via Pietro Giuria 1, I-10125 Torino, Italy}
\vskip 1.0cm
{\bf Abstract}
\end{center}
We show that any pair of classical solutions of open string field
theory can be related by a formal gauge transformation defined by a gauge 
parameter $U$ without an inverse. We investigate how this observation can be 
used to construct new solutions. We find that a choice of 
gauge parameter consistently generates a new solution only if the BRST 
charge maps the image of $U$ into itself. When this occurs,
we argue that $U$ naturally defines a star algebra projector which 
describes a surface of string connecting the boundary conformal field theories
of the classical solutions related by $U$. We also note that singular gauge 
transformations give the solution space of open string field theory the 
structure of a category, and we comment on the physical interpretation of 
this observation.

\noindent

\noindent
\medskip

\end{titlepage}

\newpage

\tableofcontents

\baselineskip=18pt

\section{Introduction}

Classical solutions in Chern-Simons-like theories \cite{Jones,Witten} are 
given by flat connections. The easiest way to find a flat connection is to 
assume that the solution can be written in the form
\begin{equation}A = U^{-1}d U.\label{eq:CSgauge}\end{equation}
Naively this implies that $A$ is pure gauge. However, 
the solution can be nontrivial if, for some reason, $U$ is not an acceptable
gauge transformation. This can happen in a couple of ways. One way is if
the gauge transformation $U$ does not belong to the space of fields used to 
define the theory. For example, in Chern-Simons theory on a 3-manifold 
$\mathcal{M}$, we can construct a Wilson line 
wrapping a noncontractible cycle using a gauge parameter $U$ which lives 
on the universal cover of $\mathcal{M}$. The string field theory analogue 
of this is to 
construct $U$ using an insertion of a boundary condition changing operator, 
so that $U$ is not a state within a single boundary 
conformal field theory.\footnote{For the Wilson line deformation on a circle, 
the analogy between string field theory and Chern-Simons theory is 
direct: The boundary condition changing operator for the Wilson
line deformation is $e^{\pm iA X}$, which is a vertex operator in 
the boundary conformal field theory whose target space is the 
universal cover of the circle.} 
This idea is a starting point for the construction
of analytic solutions for marginal deformations in open string field
theory \cite{FK,KO,FKsuper,KOsuper}.

However, $U$ can fail to be a gauge transformation for another reason: $U$ 
might not have an inverse. This is the sense that Schnabl's solution for the 
tachyon vacuum is ``almost'' pure gauge \cite{Schnabl,Okawa}. This suggests an 
appealingly simple strategy for discovering new solutions: Make an educated 
guess for $U$, and then solve the linear 
inhomogeneous equation $Q U = U\Phi$ for $\Phi$. This strategy has lead to 
interesting proposals for the tachyon lump \cite{Ian,BMT} and multiple 
branes solutions \cite{multiple,multiple2}, but unfortunately these 
constructions are singular and it is not known how to make them consistent 
with the equations of motion \cite{multiple2,lumps,multi_bdry,Hata}. The 
problem can be traced to the fact that because $U$ is not invertible, in 
general $U^{-1}QU$ is not well-defined. 

In this paper we attempt to confront this issue. First we show that 
any pair of solutions in open string field theory can be related by a 
gauge transformation of the form \eq{CSgauge}, with the understanding that $U$ 
might not be invertible. We call this a {\it left gauge 
transformation}. We then observe that the expression $U^{-1}(Q+\Phi_1)U$ 
only makes sense if $(Q+\Phi_1)U$ is equal to $U$ times something. This 
imposes a nontrivial constraint on the possible $U$s which can be used
to define new solutions. To phrase this condition in a more useful way, 
we assume the existence of a star algebra projector $X^\infty$ which projects
onto the left and right kernel of $U$. Then a consistent left gauge 
transformation must satisfy the constraint:
\begin{equation}X^\infty Q_{\Phi_1} U = 0.\label{eq:weak0}\end{equation}
where $Q_{\Phi_1}$ is the kinetic operator around the reference solution. 
Our understanding of this equation is formal, and we will not attempt to 
solve it to find new solutions in this paper. However, we show that, under 
a few assumptions, it is nontrivially satisfied for all analytic solutions 
we have studied, and it is violated for the multibrane and lump solutions
of \cite{Ian,BMT,multiple,multiple2}, which are known to encounter 
difficulties. One interesting byproduct of 
our analysis concerns the projector $X^\infty$. By formal arguments and 
examples, we find that $X^\infty$ is a projector-like state representing 
a surface of open string connecting the boundary conformal field theories of 
the classical solutions related by $U$. Accordingly, we call $X^\infty$ the
{\it boundary condition changing projector}. The boundary condition changing
projector is important not only because of the consistency condition 
\eq{weak0}, but also because it determines the structure of the 
{\it phantom term} needed to precisely define the solution in pure gauge 
form \cite{Schnabl,Okawa,FKsol,RZO,SSF2,Erler,simple,exotic}.

This paper is organized as follows. In section \ref{sec:formalism} we 
develop the formalism assuming that the string field algebra can be usefully 
modeled as a (preferably finite dimensional) algebra of operators acting on 
some vector space. We show how to relate any two 
classical solutions by a left gauge transformation, state the 
assumptions needed in order to define the boundary condition changing 
projector, and state two conditions---the {\it strong} and {\it weak} 
\integrability conditions---which every singular gauge transformation should 
satisfy in order to generate a new solution. Computing the BRST 
variation of $X^\infty$, we motivate a physical interpretation of the 
boundary condition changing projector in terms of a stretched string 
connecting two boundary conformal field theories.  We comment on the relation 
between the boundary condition changing projector and the characteristic 
projector defined by Ellwood \cite{Ian}. Finally we demonstrate the formalism 
using a finite dimensional toy model of vacuum string field theory. 
In section \ref{sec:category}, we observe that left gauge transformations 
can be interpreted as the morphisms of a category whose objects are classical
solutions. We explain how this structure can be naturally related to a 
picture of open strings ending on D-branes. In 
section \ref{sec:examples} we apply the formalism to some known analytic 
solutions in string field theory, finding that, with some assumptions, it does
a pretty good job at explaining why some candidate formal gauge 
transformations define solutions, while others do not. We also work out 
explicit examples of the boundary condition changing projector and describe 
how it implements the change of boundary condition. We end with some 
discussion.

\section{Formalism}
\label{sec:formalism}

To set up the formalism, we consider a ``model'' of the open string star 
algebra consisting of three ingredients:
\begin{description}
\item {\bf 1)} An associative algebra $\mathcal{A}$ 
with an integer grading which we call ghost number, and a $\mathbb{Z}_2$ 
grading corresponding to Grassmann parity. Grassmann parity is identified 
with ghost number mod $\mathbb{Z}_2$.  
\item {\bf 2)} A nilpotent, Grassmann odd and ghost number 1 
derivation of $\mathcal{A}$ which we call 
the BRST operator $Q$.
\item {\bf 3)} A representation of $\mathcal{A}$ as an algebra of operators 
acting on some vector space, $\D$.
\end{description}
In string field theory, $\D$ might be identified
with the space of half-string wavefunctionals \cite{RSZsplit,Gross}, but at 
present it is difficult to say much concrete about this. We will not attempt 
to be rigorous about the precise analytic definition of the operator algebra 
$\A$ and its topology; unless otherwise mentioned, we will effectively assume 
$\D$ is a finite dimensional vector space. This means that, for string field 
theory purposes, our discussion will be formal. Its relevance should be 
justified by examples, as discussed in section \ref{sec:examples}.

Note that ingredients {\bf 1)-3)} are not specific to string field 
theory, but can be realized in Chern-Simons \cite{Jones} or 
noncommutative Chern-Simons theory \cite{Susskind,Gross-Periwal}. However,
as we will see the formalism is less interesting in these examples due 
to the absence of fields with negative ghost number.

\subsection{Left Gauge Transformations}

Consider two solutions $\Phi_1$ and $\Phi_2$ related by a finite gauge 
transformation:
\begin{equation} \tilde{U}(Q+\Phi_1)U = \Phi_2,\end{equation}
where $\tilde{U}U=U\tilde{U}=1$. Multiplying this equation by $U$ from the 
left, we find the relation 
\begin{equation}(Q+\Phi_1)U = U\Phi_2,\label{eq:left}\end{equation}
and multiplying by $\tilde{U}$ from the right gives the equation
\begin{equation}-Q\tilde{U} +\tilde{U}\Phi_1=\Phi_2 \tilde{U} .\label{eq:right}
\end{equation}
We will call $U$ satisfying \eq{left} a {\it left gauge transformation} from
$\Phi_1$ to $\Phi_2$, and $\tilde{U}$ satisfying \eq{right} a 
{\it right gauge transformation} from $\Phi_1$ to $\Phi_2$. A 
left gauge transformation from $\Phi_1$ to $\Phi_2$ is also a right gauge 
transformation, in reverse, from $\Phi_2$ to $\Phi_1$. Note that these
definitions are meaningful even when $U$ or $\tilde{U}$ are not invertible, 
in which case $\Phi_1$ and $\Phi_2$ may not be gauge equivalent solutions. 
If $U$ or $\tilde{U}$ is invertible, then we will call it a 
{\it proper gauge transformation}, and, if not, {\it singular gauge 
transformation}. 

A short explanation of terminology: When discussing left gauge 
transformations, we will think of operators in the algebra $\mathcal{A}$ 
as acting {\it from the left} on the representation space $\D$. For right 
gauge transformations, it turns out to be more natural to treat operators as 
acting {\it from the right} on the dual space $\D'$. In the following 
development we will focus on left gauge transformations. The story for 
right gauge transformations is simply a ``mirror image.''

It is convenient to introduce the operator
\begin{equation}Q_{\Phi_1\Phi_2}A \equiv QA + \Phi_1 A
+(-1)^{A} A\Phi_2.
\end{equation}
This is nilpotent, but not a derivation. However it satisfies a 
modified Leibniz rule,
\begin{eqnarray}
Q_{\Phi_1\Phi_3}(AB) \lineup = (Q_{\Phi_1\Phi_2}A)B+(-1)^A A 
(Q_{\Phi_2\Phi_3}B),\label{eq:Leibniz}
\end{eqnarray}
where $\Phi_2$ on the right hand side is any solution. Rewriting \eq{left},
we can define a left gauge transformation $U$ as a ghost number zero solution
to the equation:
\begin{equation}Q_{\Phi_1\Phi_2}U=0.\end{equation}
The obvious solution is $U=0$, but this is too trivial to be 
interesting. If the theory has a nonzero field $b$ at ghost 
number $-1$, we can find a more interesting solution
\begin{equation}U = Q_{\Phi_1\Phi_2}b .\label{eq:left_ex}\end{equation}
Since string field theory has many fields with negative ghost number, this 
implies the following:
\begin{equation}\phantom{\Bigg(}{
\mbox{\it In string field theory there is always a nonzero 
left gauge transformation} \atop \mbox{\it
connecting any pair of solutions.}\ \ \ \ \ \ \ \ \ \ \ \ 
\ \ \ \ \ \ \ \ \ \ \ \ \ \ \ \ \ \ \ \ \ \ \ \ \ \ \ \ \ 
\ \ \ \ \ \ \ \ \ \ \ }\nonumber\end{equation} 
This is not the case in Chern-Simons theory. Since there are no negative rank
forms, the construction of nonzero left gauge transformations depends on 
whether $Q_{\Phi_1\Phi_2}$ has cohomology at ghost number zero. If $\Phi_1$ 
and $\Phi_2$ are gauge equivalent, $Q_{\Phi_1\Phi_2}$ will have cohomology 
by construction, but this is not guaranteed if they are not gauge equivalent. 
Therefore, the existence of nonzero singular gauge transformations is 
something particularly characteristic of string field theory. 

\subsection{Consistency Conditions and the BCC projector}

The basic question we want to ask is this: Given a solution $\Phi_1$, when 
can a field $U$ be regarded as a left gauge transformation
to another solution $\Phi_2$? From equation \eq{left} it is obvious that 
that $(Q+\Phi_1)U$ should be equal to $U$ times something. If the fields 
are linear operators acting on $\D$, this means
\begin{equation}\Box{\Img Q_{\Phi_1}U\subseteq\Img U.}
\label{eq:int1a}\end{equation}
In other words, the kinetic operator around the solution $\Phi_1$ must map 
the image of $U$ into itself. We will call this the {\it strong 
\integrability condition}. This condition implies that we can find a 
field $\Phi_2$ satisfying $(Q+\Phi_1)U=U\Phi_2$, but it does not guarantee
that $\Phi_2$ is a solution. However, $\Phi_2$ at least satisfies
\begin{equation}U(Q\Phi_2+\Phi_2^2)=0,\end{equation}
so it is a solution up to the kernel of $U$. 

\begin{figure}
\begin{center}
\resizebox{5.8in}{1.4in}{\includegraphics{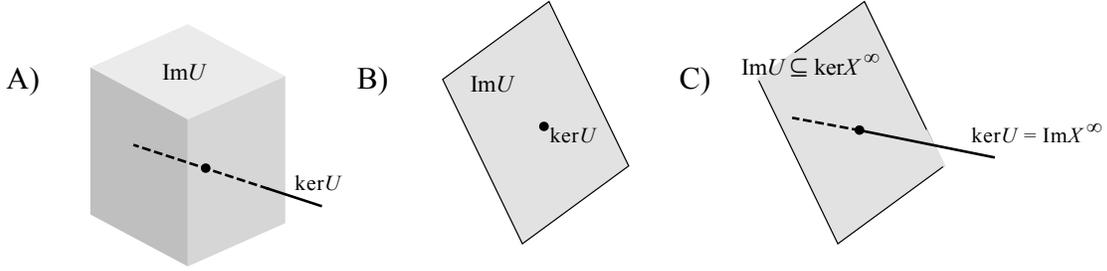}}
\end{center}
\caption{\label{fig:ker} We assume that the image and kernel of $U$ are 
linearly independent and span the whole space, as shown above in case 
C). In cases A) and B) the image and kernel of $U$ are not 
linearly independent or do not span the whole space. This would happen if 
$U$ was a non-unitary isometry.}\end{figure}

While the strong \integrability condition \eq{int1a} is true, it is not very 
helpful since checking it seems to require that we already know whether 
$\Phi_2$ exists. So let's derive a different condition which can be more 
useful. To do this, we will make an assumption about $U$:
\begin{equation}\mbox{\it The kernel and image of U 
are linearly independent and span all of}\ \D.\label{eq:ass}\end{equation}
See figure \ref{fig:ker}. This assumption is motivated by 
string field theory examples, but it is also a generic property of operators 
in finite dimensions except in degenerate cases. 
We should mention, however, that this assumption excludes the possibility 
that $U$ could be a non-unitary isometry.\footnote{A prototypical example is 
the forward shift operator $U = \sum_{n=0}^\infty |n+1\rangle\langle n|$, 
which satisfies $U^\dag U=1$, but $UU^\dag = 1-|0\rangle\langle 0|$.} 
Non-unitary isometries play an important role in the construction of 
solutions in noncommutative field theories \cite{partial}, and have been 
speculated to be important in string field theory as well \cite{partialSFT}, 
though this remains to be seen. However, suppose that the assumption 
\eq{ass} holds. Then we can uniquely 
define a projector $X^\infty$ with the property that it projects onto the 
kernel of $U$, and also annihilates states proportional to 
$U$:\footnote{In finite dimensions, the assumption \eq{ass} implies that 
$\Img U =\ker X^\infty$. In infinite dimensions, we allow for the 
possibility that the image of $U$ might only be dense in the kernel of 
$X^\infty$.}
\begin{equation}\Box{\ \ \ker U=\Img X^\infty, \ \ \ \ \ \ \ 
\Img U\subseteq\ker X^\infty.\ \ }
\label{eq:BCCfacts}\end{equation}
For reasons to be explained later, we will call $X^\infty$ the 
{\it boundary condition changing projector} (or BCC projector for short). 
We can compute the BCC projector from $U$ in many ways. For example, if we 
define 
\begin{equation}U\equiv 1-X.\end{equation}
Then if the limit exists, the infinite power of $X$ converges to the BCC 
projector:
\begin{equation}X^\infty=\lim_{N\to\infty}X^N.\end{equation}
Another useful formula for the BCC projector is the limit
\begin{equation}X^\infty =\lim_{\eps\to0^+}\frac{\eps}{\eps+ U}.
\label{eq:Ueps}\end{equation}
Often this expression is easier to work with, and converges in many cases when 
$\lim_{N\to\infty}X^N$ does not.\footnote{If $X$ is a diagonalizable matrix, 
$\lim_{N\to\infty}X^N$ converges only if its eigenvalues are equal to one or 
strictly less than one in absolute value. However, 
$\lim_{\eps\to0^+}\frac{\eps}{\eps+ U}$, always converges as 
long as $U$ does not have a continuous spectrum of negative eigenvalues in the 
neighborhood of $0$.}  Provided the BCC projector exists, the strong 
\integrability condition implies   
\begin{equation}\Img Q_{\Phi_1}U\subseteq\ker X^\infty,\label{eq:int2a}
\end{equation}
or, equivalently
\begin{equation}\Box{X^\infty Q_{\Phi_1}U=0.} \label{eq:int2aa}\end{equation}
We call this the {\it weak \integrability condition}. It is weaker than 
\eq{int1a} since (in infinite dimensions) a state can be annihilated by 
$X^\infty$ without being proportional to $U$. We will encounter an example of
this in section \ref{subsec:multiple}. However, unlike \eq{int1a}, the weak 
\integrability condition is a nontrivial constraint on $U$ which 
we can check without {\it a priori} knowledge of the existence of a solution 
$\Phi_2$.\footnote{Note that we can write a more general form of the weak 
\integrability condition,
\begin{equation}X^\infty Q_{\Phi_1\Psi}U = 0,\end{equation} 
where $\Psi$ is any classical solution. 
If we choose $\Psi=\Phi_2$ to be the target solution of the left gauge 
transformation $U$, then $Q_{\Phi_1\Phi_2}U=0$ and the weak \integrability 
condition is satisfied identically. So for a consistent left gauge 
transformation, the weak \integrability condition in any form should follow 
trivially from the equation $X^\infty U=0$.} 

Suppose we have established that $U$ satisfies the strong and weak 
\integrability conditions. How do we use $U$ to construct a new solution? 
Formally we would like to write $\Phi_2= U^{-1}(Q+\Phi_1)U$, but since $U$ 
is generally not invertible we should be more precise. Even when $U$ is not 
invertible, on the restricted domain $\D/\ker U$ we can define the inverse: 
\begin{equation}
U^{-1}:\ \Img\, U\rightarrow \D/\ker U.
\label{eq:Uinv}\end{equation}
This represents an equivalence class of operators from $\Img U$ into $\D$. 
Suppose we choose a representative of this equivalence class
\begin{equation}(U^{-1})':\Img U\to \D.\end{equation}
Then we can write $\Phi_2 = (U^{-1})'(Q+\Phi_1)U$ up to some arbitrary field
in the kernel of $U$. But the kernel of $U$ consists precisely of 
states proportional to the BCC projector. Therefore, 
\begin{equation}\Phi_2 = (U^{-1})'(Q+\Phi_1)U+X^\infty\Phi',\label{eq:phantom}
\end{equation}
where $\Phi'$ is a ghost number 1 field. The last
term is precisely the {\it phantom term} known from 
studies of analytic solutions in open string field theory 
\cite{Schnabl,Okawa,FKsol,RZO,SSF2,Erler,simple,exotic}. Unfortunately, 
the phantom term cannot be determined from knowledge of the reference 
solution $\Phi_1$ or the gauge parameter $U$ alone; it requires new 
input. In principle, it's possible that there is no consistent choice of 
phantom term which produces a solution. However, we should mention that 
the formula \eq{phantom} does not completely capture what happens in string 
field theory. In string field theory, the phantom term 
only appears in the context of a precise regularization of the solution, 
whereas in \eq{phantom} there is no regularization. This is 
because in our naive considerations we have assumed that the BCC projector 
is a well-defined object in the algebra of operators acting on $\D$. This is 
certainly not the case in string field theory, where 
projectors are singular states whose star products are generally afflicted 
with associativity anomalies \cite{Ian,cohomology}.
This implies that acceptable solutions in string field theory 
should belong to a ``well-behaved'' subspace of states where projectors are 
excluded. But since the phantom term is itself proportional to a projector, 
this requires that the phantom term must be chosen so as to cancel 
projector-like states arising from the first term in \eq{phantom}. This is 
what happens for Schnabl's solution, as the sliver state is needed to cancel 
a sliver-like contribution from the sum over derivatives of wedge 
states \cite{Schnabl}. Therefore, it is possible that the phantom term is
uniquely fixed in string field theory by considerations of regularity. 

\subsection{Physical Interpretation of the BCC Projector}
\label{subsec:bcc}

We would like to motivate a physical interpretation for the boundary condition
changing projector. For this purpose we compute
\begin{equation}QX^\infty,\end{equation}
since the BRST operator will act as a kind of probe of the internal structure
of the projector. Our derivation will turn out to be formal for string 
field theory purposes, which is most likely related to singularities of 
the BCC projector caused by the shift in boundary condition 
at the midpoint (see section \ref{subsec:KOS}). Nevertheless, the
computation of $QX^\infty$ has an important physical interpretation.

To start, we take the weak \integrability condition \eq{int2aa} and 
subtract $Q(X^\infty U)=0$ to find
\begin{equation}(QX^\infty-X^\infty \Phi_1)U=0.\end{equation}
This implies that the factor in parentheses must be proportional to 
$X^\infty$:
\begin{equation}QX^\infty-X^\infty \Phi_1 =\Pi X^\infty.\label{eq:st1}
\end{equation}
To calculate $\Pi$, multiply by $U$ from the left:
\begin{equation}UQX^\infty=U\Pi X^\infty.\end{equation}
Using $Q(UX^\infty)=0$ this becomes
\begin{equation}(QU) X^\infty =- U\Pi X^\infty.\end{equation}
Now assume that $U$ is a left gauge transformation from the solution 
$\Phi_1$ to the solution $\Phi_2$. Then $QU = U\Phi_2 - \Phi_1 U$ and
\begin{equation}U\Phi_2 X^\infty = -U\Pi X^\infty .\end{equation}
Then without loss of generality we can assume $\Pi$ takes the form 
\begin{equation}\Pi = -\Phi_2 - X^\infty M .\end{equation}
Plugging in to \eq{st1} we find:
\begin{equation}QX^\infty + \Phi_2 X^\infty +X^\infty M X^\infty- 
X^\infty\Phi_1 =0 .\label{eq:st2}\end{equation}
To determine $M$, multiply \eq{st2} by $X^\infty$ from
the left and from the right
\begin{equation}X^\infty (Q X^\infty) X^\infty 
+X^\infty(M+\Phi_2-\Phi_1)X^\infty =0.\label{eq:st3}\end{equation}
The first term in this equation is zero, as can be seen from the following 
manipulation: 
\begin{eqnarray}Q X^\infty \lineup = Q(X^\infty X^\infty X^\infty )\nonumber\\
\lineup = (QX^\infty)X^\infty X^\infty +X^\infty (Q X^\infty) X^\infty +
X^\infty X^\infty (Q X^\infty)\nonumber\\
\lineup = (QX^\infty)X^\infty  +X^\infty (Q X^\infty) X^\infty +
X^\infty (Q X^\infty)\nonumber\\
\lineup = Q( X^\infty X^\infty)  +X^\infty (Q X^\infty) X^\infty\nonumber\\
\lineup = Q X^\infty  +X^\infty (Q X^\infty) X^\infty.
\end{eqnarray}
Therefore \eq{st3} determines $M$, and the final result is 
\begin{equation}
\Box{QX^\infty + \Phi_2 X^\infty +X^\infty(\Phi_1-\Phi_2)X^\infty- 
X^\infty\Phi_1 =0} .\label{eq:BRST}\end{equation}

To motivate our interpretation of this equation, consider a wedge 
state with boundary conditions deformed by a (nonsingular) marginal current 
$V$ \cite{bcc}:
\begin{equation}e^{-(K+V)} = \sigma_{01}\Omega\,\sigma_{10},\end{equation}
where $\sigma_{01}$ is a boundary condition changing operator (BCC operator) 
which shifts from the reference boundary conformal field theory to 
the marginally deformed boundary conformal field theory, and $\sigma_{10}$ 
shifts back. Taking the BRST variation of this equation gives
\begin{equation}-cV e^{-(K+V)} +e^{-(K+V)}cV=
(Q\sigma_{01})\Omega\,\sigma_{10}+\sigma_{01}\Omega (Q\sigma_{10}).
\end{equation}
Thus we can informally identify,
\begin{equation}cV\sim Q\sigma_{10}.\end{equation}
Now note that $cV$ is a solution to the string field theory equations of 
motion. This suggests a general interpretation: A solution in open string 
field theory corresponds, from the worldsheet perspective, to the BRST 
variation of a BCC operator.

\begin{figure}
\begin{center}
\resizebox{2.8in}{2in}{\includegraphics{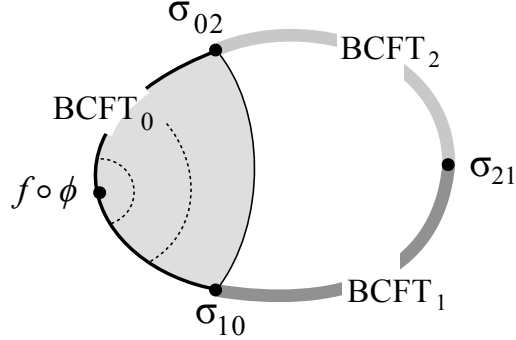}}
\end{center}
\caption{\label{fig:BCC} The BCC projector should structurally look like a 
surface of stretched string probed from a reference boundary conformal field
theory. As shown above, such a surface should have an insertion 
of the boundary condition changing operator $\sigma_{21}$ between the BCFTs
at the two endpoints. It should also have boundary condition changing 
operators $\sigma_{02}$ and $\sigma_{10}$ to probe with a test state.}
\end{figure}

\begin{figure}
\begin{center}
\resizebox{6.2in}{2in}{\includegraphics{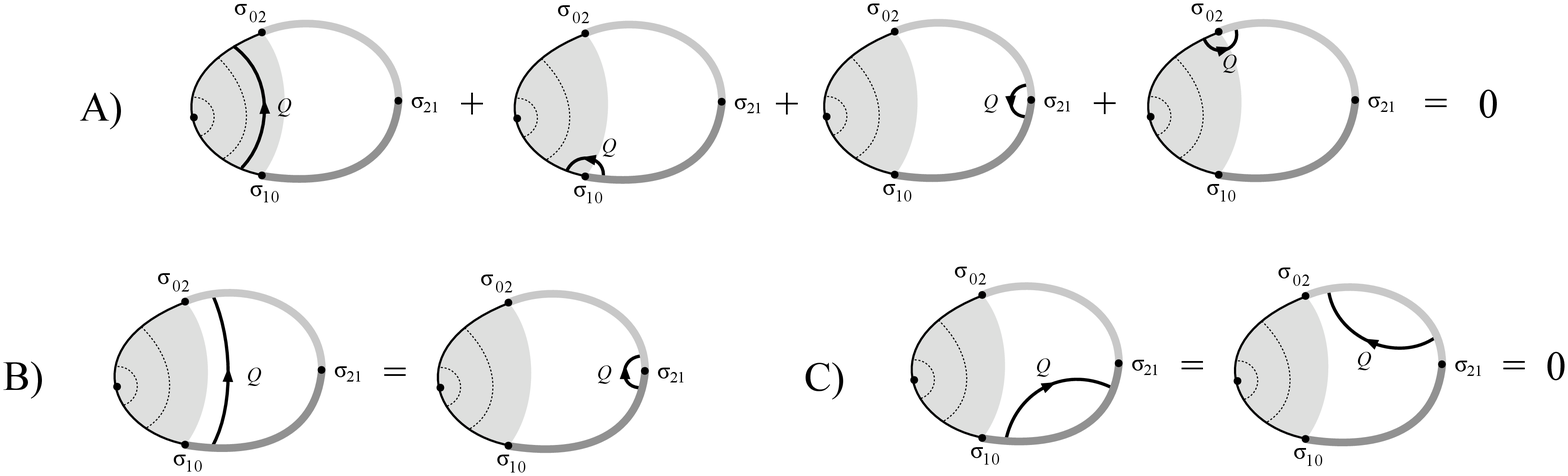}}
\end{center}
\caption{\label{fig:BCC2} Figure A) demonstrates the analogy of 
equation \eq{BRST}, B) demonstrates the analogy of \eq{BRST2} and 
C) the analogy of \eq{BRST3}.}
\end{figure}

Then an interpretation of the identity \eq{BRST} immediately presents itself:
$X^\infty$ is a star algebra projector representing a stretched string 
connecting the boundary conformal field theories of  $\Phi_2$ and $\Phi_1$. 
To see this, suppose we represent a stretched string as a 
surface state with an insertion of a boundary condition changing operator 
$\sigma_{21}$ between the BCFTs of $\Phi_2$ and $\Phi_1$. In order to probe 
this surface with a test state, we need to insert two other BCC operators 
$\sigma_{02}$ and $\sigma_{10}$ on either side of $\sigma_{21}$ to match the 
boundary condition of the reference BCFT (See figure \ref{fig:BCC}).
Now if we compute the BRST variation of this object, we find a direct correspondence with the 
BRST variation of $X^\infty$: $\Phi_2$ corresponds to the the BRST variation 
of $\sigma_{02}$; $\Phi_1-\Phi_2$ corresponds to the BRST variation of 
$\sigma_{21}$; and $\Phi_1$ corresponds the BRST variation of $\sigma_{10}$ 
(See figure \ref{fig:BCC2} A). 
The projector $X^\infty$ may not appear quite as simple as figure 
\ref{fig:BCC}, since the way that string field theory represents worldsheet 
boundary conditions can be indirect. However, the shift in boundary condition 
inside $X^\infty$ is remarkably clear in the examples we have studied. 
Therefore, we call $X^\infty$ the {\it boundary condition changing projector}.

The identity \eq{BRST} can be written in a few other forms which help 
illuminate the interpretation of $X^\infty$. For example,
\begin{equation}Q_{\Phi_2\Phi_1}X^\infty = 
X^\infty (Q_{\Phi_2\Phi_1}1) X^\infty\label{eq:BRST2}\end{equation}
is analogous to the statement that the BRST variation of a surface 
of stretched string only receives contribution from the boundary condition 
changing operator between the open string endpoints. (See figure 
\ref{fig:BCC2} B). Multiplying  \eq{BRST2} by $X^\infty$ on either side, 
we also find
\begin{equation}(Q_{\Phi_2}X^\infty)X^\infty = X^\infty(Q_{\Phi_1}X^\infty)
 = 0.\label{eq:BRST3}\end{equation}
This corresponds to the statement that the boundary conditions are BRST 
invariant separately on each endpoint of the open string. (See figure 
\ref{fig:BCC2} C).

When calculating the BRST variation of $X^\infty$ we assumed that 
all operators could be treated like finite dimensional matrices. However, 
in string field theory the double-projector term 
$X^\infty(\Phi_1-\Phi_2)X^\infty$ in \eq{BRST} is problematic, 
since star products of projector-like states are not in general 
well-defined. Therefore in 
string field theory equation \eq{BRST} should be understood in the context 
of some regularization. Let us present two regularizations. 
The first represents the BCC projector as the limit
\begin{equation}X^\infty = \lim_{\eps\to0^+}\frac{\eps}{\eps+ U}.
\end{equation}
Assuming $U$ is a left gauge transformation from $\Phi_1$ to $\Phi_2$, 
we can easily calculate
\begin{eqnarray}Q\frac{\eps}{\eps+ U} \lineup = 
-\frac{\eps}{\eps+ U} QU\frac{1}{\eps+ U}\nonumber\\
\lineup = -\frac{\eps}{\eps+ U}
\Big[(U+\eps)\Phi_2 - \Phi_1(U+\eps)+\eps(\Phi_1-\Phi_2)\Big]
\frac{1}{\eps+ U}\nonumber\\
\lineup = -\Phi_2\frac{\eps}{\eps+ U} + \frac{\eps}{\eps+ U}\Phi_1 
-\frac{\eps}{\eps+ U}(\Phi_1-\Phi_2)\frac{\eps}{\eps+ U}.
\end{eqnarray}
Therefore
\begin{equation}Q\frac{\eps}{\eps+ U}+\Phi_2\frac{\eps}{\eps+ U}
+\frac{\eps}{\eps+ U}(\Phi_1-\Phi_2)\frac{\eps}{\eps+ U}-
\frac{\eps}{\eps+ U}\Phi_1=0.\end{equation}
Note that this reproduces the basic form of \eq{BRST} even before taking
the $\eps\to 0$ limit. The second regularization represents the BCC 
projector as an infinite power of $X$:
\begin{equation}X^\infty = \lim_{N\to\infty}X^{N}.\end{equation}
With some algebra one can prove the identity
\begin{equation}Q X^{2N} +\Phi_2 X^{2N-1} +X^N(\Phi_1-\Phi_2)X^N 
- X^{2N-1}\Phi_1 = \mathcal{R}(N),\label{eq:BRSTreg2}\end{equation}
where the remainder $\mathcal{R}(N)$ is the expression 
\begin{eqnarray}\mathcal{R}(N)\lineup =\left[\Phi_1 X^{N-1}+X
\sum_{k=0}^{N-2} X^{N-1-k}(\Phi_1-\Phi_2)X^k\right]UX^N\nonumber\\
\lineup\ \ \ +UX^N\left[-X^{N-1}\Phi_2 
+\sum_{k=0}^{N-2}X^{N-2-k}(\Phi_1-\Phi_2)X^{k+1}\right].\end{eqnarray}
Equation \eq{BRSTreg2} reproduces \eq{BRST} if $\mathcal{R}(N)$ vanishes  
in the $N\to\infty$ limit. The remainder does vanish if 
$X^N(\Phi_1-\Phi_2)X^N$ is finite for large $N$.  

\subsection{BCC Projector vs. Characteristic Projector}

Let us explain the relation between the BCC projector and 
the characteristic projector introduced by Ellwood \cite{Ian}. The 
characteristic projector is defined given an arbitrary solution $\Phi$ 
together with a reference solution $\Psi$ and homotopy operator $A$ satisfying
\begin{equation}Q_{\Psi}A=1,\ \ \ \ \ A^2=0.\end{equation}
This implies that the kinetic operator around $\Psi$ supports no cohomology, 
and therefore $\Psi$ can be interpreted as the tachyon 
vacuum \cite{cohomology,cohomology1}. The characteristic projector is 
defined\footnote{We use the bracket $[\,,]$ to denote the graded commutator.}
\begin{equation}P \equiv \lim_{N\to\infty}
(-[A,\Phi-\Psi])^N.\label{eq:Ian}\end{equation}
We claim that the characteristic projector is the BCC projector for a singular
gauge transformation from a solution to itself. To see this, note that 
\begin{eqnarray}U_1 \lineup = Q_{\Phi\Psi}A\nonumber\\
\lineup =1+(\Phi-\Psi)A\end{eqnarray}
is a left gauge transformation from $\Phi$ to the tachyon vacuum, and 
\begin{eqnarray}U_2 \lineup = Q_{\Psi\Phi}A\nonumber\\
\lineup = 1+A(\Phi-\Psi)\end{eqnarray}
is a left gauge transformation from the tachyon vacuum to $\Phi$. 
Therefore the product,
\begin{equation}U=U_1U_2=1+[A,\Phi-\Psi],\label{eq:PhiPsiPhi}\end{equation}
is a left gauge transformation from $\Phi$ to itself. To find the BCC 
projector we take the infinite power of
\begin{equation}X = -[A,\Phi-\Psi].\end{equation}
This is just the characteristic projector. In  \cite{Ian} it was conjectured, 
and demonstrated in examples, that the characteristic projector is a 
sliver-like state representing the boundary conditions of $\Phi$ deep in its 
interior. The BCC projector should represent a change of 
boundary condition between two BCFTs. But in this case the source and target 
solutions are the same, so we only see the boundary conditions of a single 
BCFT, consistent with Ellwood's interpretation of the characteristic projector.

One of the main insights of \cite{Ian}, which was an inspiration for the 
current work, was that singular gauge transformations could be thought of as 
possessing a kernel which could be described with a star algebra projector. 
However, the treatment of singular gauge transformations presented here 
differs substantially from \cite{Ian}. A central assumption of \cite{Ian}, 
which we do not follow, is that the characteristic projector is annihilated 
by the homotopy operator:
\begin{equation}AP = PA = 0.\label{eq:Ell_ass}\end{equation}
This equation is true in known examples in string field theory, provided 
one is only concerned with string fields as they are defined in the Fock 
space expansion. However, this assumption leads to a number of apparent 
inconsistencies. For example, the BCC projector for the left gauge 
transformation $U_2$ from the tachyon vacuum to $\Phi$ can be written
\begin{equation}X^\infty = -PA(\Phi-\Psi)\end{equation}
Assuming \eq{Ell_ass} this means $X^\infty=0$, which implies that $U_2$ 
has no kernel and all solutions should be gauge 
equivalent to the tachyon vacuum. On the other hand, 
\eq{Ell_ass} also implies
\begin{equation}U_2 P = 0\end{equation}
which contradicts what we just proved, i.e. that $U_2$ has no kernel. 
The contradiction comes because \eq{Ell_ass} implies an associativity 
anomaly \cite{Ian}
\begin{equation}P(U_2 P) = 0 \neq P = (P U_2)P\end{equation}
This means that \eq{Ell_ass} can never be true in the type of 
matrix-like model of the string field algebra we have been assuming. 
Therefore in our approach \eq{Ell_ass} is false, which means that
we require a stronger notion of equality than the Fock space expansion of the
string field. Indeed, we believe that this is physically necessary since 
\eq{Ell_ass} implies that the phantom term for Schnabl's solution vanishes, 
which misses the nontrivial contribution the phantom term makes to gauge 
invariant observables. We will return to this issue in section 
\ref{subsec:Schnabl}.

\subsection{Toy Model}
\label{subsec:toy}

Before considering string field theory, it is helpful to see how the 
formalism is supposed to work in a finite dimensional toy model. Suppose 
the open string star algebra is just the Clifford algebra 
generated by two elements $b,c$ satisfying
\begin{equation}[b,c]=1,\ \ \ \ \ \ b^2=c^2=0.\label{eq:toy_alg}\end{equation}
These elements are Grassmann odd and have the obvious ghost number.
The algebra allows a 2-dimensional representation $\D$ in terms of
Pauli matrices. We define the BRST operator to be 
\begin{equation}Q=[c,\cdot].\end{equation}
The cohomology is empty. Therefore, our toy model can be thought of as a 
simplified version of vacuum string field theory \cite{VSFT}.

The equation of motion,
\begin{equation}Q\Phi+\Phi^2=0,\end{equation}
is easy to solve:
\begin{equation}\Phi = -\lambda c,\ \ \ \ \lambda\in \mathbb{R}.\end{equation}
What is less obvious is whether any of these solutions is physically 
nontrivial. We can construct a left gauge 
transformation relating the tachyon vacuum $\Phi=0$ and the general solution 
$\Phi=-\lambda c$:
\begin{eqnarray}U \lineup = Q_{0,-\lambda c}b\nonumber\\
\lineup = 1-\lambda bc.\end{eqnarray}
Computing the BCC projector with \eq{Ueps},\footnote{We can also compute the 
limit $\lim_{N\to\infty}X^N$, but this diverges if $|\lambda|> 1$. This 
does not indicate the absence of a BCC projector for $\lambda>1$.} we find:
\begin{equation}X^\infty 
=\lim_{\eps\to0}\left(\eps+\frac{\eps\lambda}{1-\lambda+\eps\lambda}bc\right).
\end{equation}
The limit vanishes in all cases except $\lambda=1$, where the BCC projector 
becomes
\begin{equation}X^\infty =bc, \ \ \ \ \ \ (\lambda=1).\end{equation}
Therefore our toy model has only one nontrivial 
solution:
\begin{equation}\Phi=-c.\label{eq:toy_sol}\end{equation}
Note that the projector appears discontinuously at $\lambda=1$, and
not in the $\lambda\to 1$ limit. This is reminiscent of how Schnabl's 
solution formally appears to be a limit of pure gauge solutions as 
$\lambda\to 1$, but at $\lambda=1$ there is a physical discontinuity which 
brings the solution to the tachyon vacuum. A similar discontinuity appears for
the pure gauge and tachyon vacuum solutions of Takahashi and Tanimoto 
\cite{TT,KT,KT2}. In this toy model, what 
makes the solution $\Phi=-c$ different from the others is that it 
supports cohomology. In fact, the shifted kinetic operator vanishes 
identically:
\begin{equation}Q-[c,\cdot]=0,\end{equation}
and therefore any nonzero state is in the cohomology. At ghost number 1 the
cohomology includes only $c$, which we can interpret as the tachyon 
of an unstable brane. Therefore the solution $\Phi=-c$ represents
a D-brane sitting on top of the tachyon vacuum.

Though the equations of motion are easy to solve in this model, 
let's try to construct the solutions indirectly using 
singular gauge transformations. Within this subalgebra, there are only two 
nonzero and noninvertible ghost number zero fields which could give an
interesting solution
(up to a trivial multiplicative factor):
\begin{eqnarray}U_1 \lineup =1- bc,\ \ \to\ \ X^\infty_1 = bc,\nonumber\\
U_2 \lineup = 1-cb,\ \ \to\ \  X_2^\infty =cb.
\end{eqnarray}
To the right of the arrow we wrote the corresponding BCC projector. 
Starting from the tachyon vacuum $\Phi=0$, the first case obviously 
corresponds to the solution $\Phi=-c$ which we have already discovered. 
What about the second case? It is easy to check that the 
weak \integrability condition is not obeyed:
\begin{equation}X_2^\infty QU_2 = (cb)Q(1-cb)=c\neq 0,\end{equation}
so there is no corresponding solution. Now, given $U_1$, how do we reconstruct
the solution $\Phi=-c$? Following \eq{phantom}, we need to define a formal 
inverse for $U_1$. It suffices to choose
\begin{equation}(U_1^{-1})'=1,\end{equation}
since this inverts $U_1$ up to the kernel of $U_1$. Then
\begin{eqnarray}\Phi \lineup = (U_1^{-1})'QU_1 +X_1^\infty \Phi'\nonumber\\
\lineup = Q(1-bc)+bc\Phi'\nonumber\\
\lineup = -c +bc\Phi'.
\end{eqnarray}
Here we are lucky that any choice of $\Phi'$ must be proportional to $c$, 
which is killed by $bc$.  Therefore the phantom term vanishes and the 
solution is uniquely determined by $U_1$. 

To see an example of a nontrivial phantom term, we have to consider a more
complicated model. Suppose that the algebra consists of Clifford algebra
generated by four elements $c_1,c_2,b_1,b_2$ satisfying
\begin{equation}[b_i,c_j]=\delta_{ij},\ \ \ \ b_i^2 = c_i^2 =0\ \ \ \ \ 
\ \ \ \ (i,j=1,2).
\end{equation}
with the obvious ghost number assignments. Taking $Q=[c_1,\cdot]$, consider
the solution
\begin{equation}\Phi=-c_1 + b_1c_1c_2.\label{eq:phant_ex}\end{equation}
This solution is nontrivial because the shifted kinetic operator supports 
cohomology. At ghost number 1 the cohomology is 1-dimensional and consists
of states proportional to $c_2$ modulo exact terms:
\begin{equation}Q_\Phi c_2 =0,\ \ \ c_2\neq Q_\Phi(\mathrm{something}).
\end{equation}
We can find a left gauge transformation from the tachyon vacuum to the solution
\eq{phant_ex}:
\begin{equation}U=Q_{0\Phi}b_1 = 1-b_1c_1.\end{equation}
The BCC projector is $X^\infty = b_1c_1$. Choosing $(U^{-1})'=1$ we can 
reconstruct the solution out of $U$:
\begin{eqnarray}\Phi \lineup = (U^{-1})'QU + X^\infty \Phi'\nonumber\\
\lineup = -c_1 +X^\infty\Phi'.\end{eqnarray}
If we want the solution we started with, apparently we must have a nonzero
phantom term:
\begin{equation}X^\infty \Phi' = (b_1c_1)c_2.\end{equation}
Note that this choice requires additional information not contained in the 
left gauge transformation. In fact, we could have chosen the phantom term
to vanish, though the resulting solution $\Phi=-c_1$ is physically different
from the solution we started with. (The spectrum of fluctuations around 
$\Phi=-c_1$ includes all ghost number 1 states, not just $c_2$). 
As mentioned before, in string 
field theory the situation may be different, since regularity may fix the 
phantom term uniquely once we have chosen $U$. 

\section{The Category of Classical Solutions}
\label{sec:category}

It is interesting to ask what happens if we generalize the gauge group of 
open string field theory to include singular gauge transformations. 
Obviously we don't have a group anymore since we don't have inverses. But it 
is not even a semi-group, since the product of two left gauge transformations
is not generally a left gauge transformation. However, it {\it is} a left 
gauge transformation if the target solution of the first left gauge 
transformation matches the source solution of the second. Let $U_{12}$ be 
a left gauge transformation from $\Phi_1$ to $\Phi_2$, and $U_{23}$ be 
a left gauge transformation from $\Phi_2$ to $\Phi_3$. Then the Leibniz 
rule \eq{Leibniz} implies
\begin{equation}Q_{\Phi_1\Phi_3}(U_{12}U_{23}) 
= (Q_{\Phi_1\Phi_2}U_{12})U_{23}+
U_{12}(Q_{\Phi_2\Phi_3}U_{23})=0,\end{equation}
so $U_{12}U_{23}$ is a left gauge transformation from $\Phi_1$ to 
$\Phi_3$. Therefore multiplication of left gauge transformations works like
the composition of maps; we can only compose two maps if the image of the 
first is contained in the domain of the second.

The structure we're describing is a {\it category}, which we call {\bf Left}. 
The objects of {\bf Left} are classical solutions, and the morphisms are left 
gauge transformations. Composition of morphisms is associative because the 
star product is associative. Each object has an identity morphism, which is
just the identity string field $U=1$.\footnote{The category description 
allows us to import some terminology: A proper gauge transformation between 
equivalent solutions is an {\it isomorphism}; The left gauge transformation 
$U=0$ is a {\it zero morphism}, and the category whose morphisms are
right gauge transformations is the {\it opposite category} from {\bf Left}.}

The category {\bf Left} is a nice description of a structure, but we would 
like to get some insight into its physical meaning. To start, note that the
operator $Q_{\Phi_1\Phi_2}$ is the kinetic operator for a stretched string 
in a $2\times2$ string field theory expanded around the classical solution 
\begin{equation}{\bf \Phi}=\left(\begin{matrix}\Phi_1 & 0 \\ 0 & \Phi_2
\end{matrix}\right).\end{equation}
Then the morphisms of {\bf Left} consist of ghost number zero states
which are closed under the action of the kinetic operator of a stretched 
string. An important subset of these morphisms
are those which are not only closed, but exact, i.e. take the form
\begin{equation}U=Q_{\Phi_1\Phi_2}b.\end{equation}
We call these {\it exact left gauge transformations}. They form an ideal 
in {\bf Left}, in the sense that the composition of any left gauge 
transformation with an exact left gauge transformation is again exact. 
As we will see, exact left gauge transformations are what make the 
category {\bf Left} interesting.

\begin{figure}
\begin{center}
\resizebox{4.5in}{1.6in}{\includegraphics{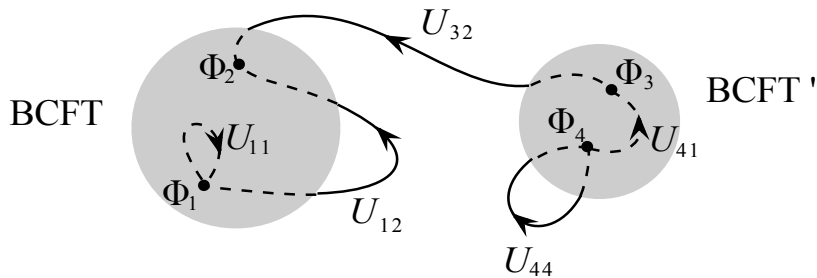}}
\end{center}
\caption{\label{fig:left} Schematic picture of the category {\bf Left}.
The points represent classical solutions, the curves represent left gauge 
transformations, and the grey circles enclose solutions describing 
the same BCFT. The curves inside the grey circles are 
proper gauge transformations, and those outside the circles are singular. 
Singular gauge transformations are often exact, and can be interpreted as 
generating surfaces of open string connecting BCFTs. Therefore the curves 
in this picture also represent open strings ending on D-branes.}
\end{figure}

Consider the exact left gauge transformation
\begin{equation}K=QB,\end{equation}
which relates the perturbative vacuum to itself. The string field $K$ 
has an important property: It is a worldsheet 
Hamiltonian, and it generates a 1-parameter family of surfaces which define 
wedge states \cite{Okawa}:
\begin{equation}\Omega^t=e^{-tK}.\end{equation}
In fact, this appears to be general: an exact left gauge transformations is 
a (generalized) BRST variation of an antighost, and therefore 
it is natural to think of it as generating a surface. If the ``wedge state'' 
$e^{-tU}$ converges in the $t\to\infty$ limit, we should get the BCC projector,
\begin{equation}X^\infty = \lim_{t\to\infty}e^{-tU},\end{equation}
which, as we have conjectured, describes a surface of stretched string 
connecting two BCFTs. Therefore
exact left gauge transformations have ``endpoints'' which are fixed to a 
source and target solution because open strings have endpoints fixed to 
corresponding D-branes. This gives a physical interpretation of the category 
{\bf Left}: The objects, up to isomorphism, represent D-branes, and the 
morphisms represent open strings connecting them.
(See figure \ref{fig:left}). This structure is reminiscent of the 
description of D-branes in terms of derived categories 
\cite{Douglas,Sharpe,AL}, and it would be interesting to explore the relation.

One aspect of this picture requires explanation. Exact left gauge
transformations are not necessarily singular, or vice-versa. Yet both should
define a BCC projector with an open string interpretation.
To explain the relation between exact and singular gauge transformations, 
we state two facts:
\begin{fact2}
The only exact left gauge transformations which are invertible relate 
the tachyon vacuum to itself.
\label{fact:inverse_exact}
\end{fact2}
\begin{proof} Suppose an exact left gauge transformation 
$U_{12}=Q_{\Phi_1\Phi_2}b$ has an inverse, $U_{12}^{-1}$.
Then
\begin{equation}1=U_{12}^{-1}U_{12} = Q_{\Phi_2\Phi_2}(U_{12}^{-1}b)=
Q_{\Phi_1\Phi_1}(bU_{12}^{-1}).\end{equation}
This implies that both solutions $\Phi_1$ and $\Phi_2$ support no cohomology 
at any ghost number, so they must both describe the tachyon vacuum.
\end{proof}

\noindent This result is consistent with the expectation that exact left 
gauge transformations generate surfaces, and, at the same time, only 
singular gauge transformations define a nonzero BCC projector. The only case 
where an exact left gauge transformation can be nonsingular is around the 
tachyon vacuum, where there is no open string surface to generate.

\begin{fact2}Any singular gauge transformation between two solutions 
is exact provided that the spectrum of stretched strings between the two 
corresponding backgrounds has no cohomology at ghost number $0$. 
\end{fact2}
\begin{proof}
 This follows immediately from the assumption that the cohomology
of $Q_{\Phi_1\Phi_2}$ should reproduce the cohomology of a stretched string 
connecting the BCFTs corresponding to the solutions $\Phi_1$ and $\Phi_2$.
\end{proof}

\noindent Generally, an open string connecting two 
D-branes will have different boundary conditions at its two endpoints, and 
therefore will have no cohomology at ghost number zero. So in the
general situation, singular gauge transformations are exact. However, let 
us give two counterexamples. Consider a string field theory with 
$2\times 2$ Chan-Paton factors and two solutions
\begin{equation}{\bf\Phi}_1 = 0, \ \ \ \ \ \ \ \ \ {\bf\Phi}_2=
\left(\begin{matrix} 0 & 0 \\ 0 & \Phi \end{matrix}\right),\end{equation}
where $\Phi$ is a solution of the string field theory defined by the 
$2$-$2$ strings. Then
\begin{equation}{\bf U}=\left(\begin{matrix}1 & 0 \\ 0 & Q_{0\Phi}b
\end{matrix}\right)\end{equation}
is a singular gauge transformation from ${\bf \Phi}_1$ to ${\bf \Phi}_2$ 
which is not exact, due to the identity string field in the $1$-$1$ component.
The kernel of ${\bf U}$ is contained solely in the $2$-$2$ component, and 
the resulting BCC projector describes a single string stretching 
from the perturbative vacuum of the second D-brane to the background 
$\Phi$ of the second D-brane. There is no surface generated for strings 
attached to the first D-brane. To give a second example, consider the 
left gauge transformation 
\begin{equation}U = 1-2\Omega,\end{equation}
which relates the perturbative vacuum to itself. The level expansion of $U$ 
starts with $-|0\rangle$, so it is not an exact left gauge transformation. 
However, the inverse $U^{-1}$ appears to be 
divergent, or at least it is not possible to express it as a superposition 
of wedge states \cite{exotic,lightning}. This problem can be understood 
from the fact that $U$, as a function of $K$, has a zero at $K=\ln 2$. 
The BCC projector of $U$ is formally
\begin{equation}X^\infty(K) = \left\{\begin{matrix}1&\ \mathrm{for}\ K=\ln 2\\
0& \ \mathrm{otherwise}\end{matrix}\right..\label{eq:weird}\end{equation}
This projector can be considered as a limiting case of the class of 
infinite-rank projectors discussed in  \cite{exotic}. It vanishes in the 
Fock space, and it does not have a known representation in terms of open 
string surfaces. We do not know whether $U$ should be considered a singular 
gauge transformation in a physically important sense, or whether defining 
its inverse is just a technical problem.

\noindent

\section{Examples}
\label{sec:examples}

In this section we demonstrate the weak \integrability condition and the 
construction of the BCC projector for some known analytic solutions. 

We will employ the algebraic notation for wedge states with insertions 
developed in  \cite{Okawa,SSF1}, following the conventions explained in 
appendix A of  \cite{simple}. Let us review the essentials. We will frequently
employ the string fields $K$ and $B$ introduced in \cite{Okawa}, 
which correspond to vertical line integral insertions of the energy-momentum 
tensor and $b$-ghost in the cylinder coordinate system. We have $QB=K$, 
$[B,K]=0$, and $B^2=0$. Exponentials of $K$ define wedge states 
\cite{RZ_wedge}:
\begin{equation}\Omega^t = e^{-t K},\end{equation}
which are star algebra powers of the $SL(2,\mathbb{R})$ vacuum 
$\Omega\equiv|0\rangle$. The infinite power of the vacuum is a projector of 
the star algebra, the sliver state $\Omega^\infty$ \cite{RZ_wedge,RSZproj}. 
We will also encounter other 
string fields $c$, $\sigma_{10}$, and so on, which correspond to insertions 
of operators on the 
open string boundary in the cylinder coordinate system. For properties
of these insertions we direct readers to the appropriate references where the
solutions are described in detail.

\subsection{Trivial case: $U=0$}

The string field $U=0$ is a left gauge transformation between any two 
solutions. The associated BCC projector is the identity string field:
\begin{equation}X^\infty=1.\end{equation}
We can interpret this as a projector where all boundary 
condition changing operators have collapsed on top of one 
another and canceled out. Both the strong and weak \integrability conditions 
are satisfied. Using \eq{phantom} we can therefore 
express $\Phi_2$ as a formal gauge transformation of $\Phi_1$:
\begin{equation}\Phi_2 = (0^{-1})'(Q+\Phi_1)0 + 1\cdot\Phi'.\label{eq:trpg}
\end{equation}
Since there are no nonzero vectors in the image of $U=0$, the operator 
$(0^{-1})'$ must vanish over its entire domain. Then $\Phi_2=\Phi'$, and 
the entire solution consists of the phantom term. Not surprisingly, 
$U=0$ does not give any information about how to construct $\Phi_2$ from 
$\Phi_1$.

\subsection{Schnabl's solution}
\label{subsec:Schnabl}

Schnabl's solution for the tachyon vacuum takes the form\footnote{We focus
on Schnabl's solution, though the discussion is similar for other tachyon 
vacuum solutions in the $KBc$ subalgebra \cite{SSF2,simple,SSF1,id1,id2}. 
It would also be interesting to understand the tachyon vacuum solution of 
Takahashi and Tanimoto \cite{TT,KT,KT2} from the perspective of this 
formalism.}
\begin{equation}\Psi = \sqrt{\Omega}c\frac{KB}{1-\Omega}c\sqrt{\Omega}.
\label{eq:Schnabl}\end{equation}
As discovered by Okawa \cite{Okawa}, Schnabl's solution
can be constructed as a left gauge transformation of the perturbative vacuum:
\begin{equation}QU = U\Psi,\end{equation}
where
\begin{equation}U = 1-\sqrt{\Omega}cB\sqrt{\Omega}.\label{eq:Okawa}
\end{equation}
Now we want to compute the boundary condition changing projector and verify 
that Okawa's $U$ satisfies the weak \integrability condition. This will 
quickly lead to some puzzles, but with a few assumptions the formalism works 
consistently with our understanding of the physics.

The BCC projector is easy to compute from $U$:
\begin{equation}X^\infty =\sqrt{\Omega}cB\Omega^\infty. \label{eq:ptv}
\end{equation}
Since the field $B$ annihilates the sliver state in the Fock space, 
the BCC projector vanishes in the Fock space. But this means that Okawa's 
$U$ should have an inverse! In fact, it {\it is} invertible in the Fock space: 
\begin{equation}U^{-1} = 1
-\sqrt{\Omega}cB\frac{1}{1-\Omega}\sqrt{\Omega},\ \ \ \ \ \ \ 
(\mathrm{formally}).\end{equation}
Expanding the factor $\frac{1}{1-\Omega}$ as a geometric series produces a 
linear divergence proportional to the sliver state, but this divergence is 
annihilated by $B$, so in total $U^{-1}$ is finite. This raises the obvious
question: Why is Schnabl's solution not pure gauge? The point is that it 
is not enough for $U$ to be invertible in the level expansion; it must 
be invertible from the perspective of the gauge invariant action,
for example when contracting the solution with itself \cite{Okawa}. 
But in this context it is no longer true that $B$ annihilates the sliver 
state. This can be seen from the fact that the phantom term for Schnabl's 
solution,
\begin{equation}\lim_{N\to\infty}\psi_N = \sqrt{\Omega}cB\Omega^\infty 
c\sqrt{\Omega},\label{eq:psiN}\end{equation}
makes a nontrivial contribution to the energy \cite{Schnabl}. Therefore, 
we must have
\begin{equation}B\Omega^\infty\neq 0.\label{eq:Bneq0}\end{equation}
Under this assumption, the inverse of $U$ is divergent and the BCC projector
is nonzero. Consistently, Schnabl's solution is not pure gauge.

\begin{figure}
\begin{center}
\resizebox{2.5in}{1.6in}{\includegraphics{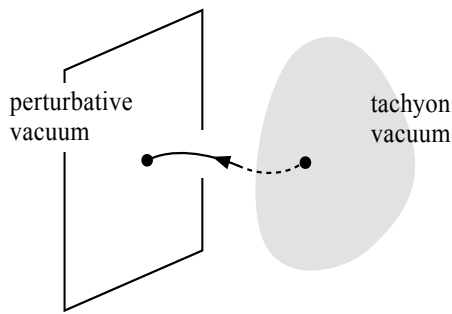}}
\end{center}
\caption{\label{fig:tv} The BCC projector for Okawa's left gauge transformation
should describe a (hypothetical) open string with an endpoint disappearing 
into the tachyon vacuum.}
\end{figure}

The above subtlety with $B\Omega^\infty$ may have a physical origin. 
Following the discussion of section \ref{subsec:bcc}, the BCC projector 
\eq{ptv} should (in principle) represent an open string connecting the tachyon 
vacuum and the perturbative vacuum, as in figure \ref{fig:tv}. But there is no
such open string; Any correlator with a boundary segment ``attached'' 
to the tachyon vacuum should vanish identically. Consistently,
the BCC projector \eq{ptv} vanishes in the Fock space. For 
other solutions we study, it will not vanish. Since the phantom 
term is proportional to the BCC projector, this also explains 
why the phantom term for Schnabl's solution vanishes in the Fock space.

Equation \eq{Bneq0} implies that the BCC projector is nontrivial, 
but to apply the weak \integrability condition we must 
also be able to assume that it is finite. This requires
\begin{equation}K\Omega^\infty= 0,\label{eq:Keq0}\end{equation}
since, if the infinite power of the $SL(2,\mathbb{R})$ vacuum
converges to a limit, we must have
\begin{equation}\lim_{t\to\infty}\frac{d}{dt}\Omega^t = 0.\end{equation}
The distinction between the string fields $B$ and $K$ with regard to the 
sliver state is subtle. We will 
proceed with the assumption \eq{Keq0} and see where it leads. 

Let us give two simple examples which help illustrate the meaning of 
equations \eq{Bneq0} and \eq{Keq0} for the purposes of the weak \integrability 
condition. Consider the equation
\begin{equation}B(1-\Omega) = KA,\label{eq:ex1}\end{equation}
which we want to solve for $A$. Since $K$ has a kernel, the solution may not
exist. To check this, multiply by the equation by sliver state:
\begin{equation}B\Omega^\infty(1-\Omega)=0.\end{equation}
The quantity $\Omega^\infty(1-\Omega)$ vanishes by \eq{Keq0}, 
so the constraint is consistently satisfied. Indeed, the solution 
$A=B\frac{1-\Omega}{K}$ is the homotopy operator for Schnabl's solution
\cite{cohomology,SSF2}, and is a well defined string field. Now, by contrast, 
consider the equation
\begin{equation}B = K H,\end{equation}
which we want to solve for $H$. This time multiplying both sides by the 
sliver gives 
\begin{equation}B\Omega^\infty\? 0,\end{equation} which contradicts 
\eq{Bneq0}. Indeed, the formal solution $H=\frac{B}{K}$ does not exist, 
since otherwise $Q\frac{B}{K}=1$, which would trivialize the cohomology of 
physical states around the perturbative vacuum. 

With this preparation, we can check the weak \integrability condition
for Okawa's $U$.
\begin{eqnarray}X^\infty QU 
\lineup = \sqrt{\Omega}cB\Omega^\infty (cKBc\sqrt{\Omega})\nonumber\\
\lineup = \sqrt{\Omega}cB(K\Omega^\infty)c\sqrt{\Omega}\nonumber\\
\lineup = 0,
\label{eq:Sch_weak}\end{eqnarray}
where in the last step we used \eq{Keq0}. This result is consistent with 
the fact that Schnabl's solution exists.

We can also consider a different left gauge transformation $\tilde{U}$ which 
maps (in the opposite direction) from Schnabl's solution to the perturbative 
vacuum:\footnote{The expression $\tilde{U}$ was also written down in 
\cite{Okawa}, where (from the current perspective) it was interpreted as a 
right gauge transformation from the perturbative vacuum to Schnabl's 
solution. This is the same as a left gauge transformation, in the opposite 
direction, from Schnabl's solution to the perturbative vacuum.}
\begin{equation}(Q+\Psi)\tilde{U} = 
\tilde{U}\cdot 0,\end{equation}
where
\begin{equation}\tilde{U}=1-\sqrt{\Omega}Bc\sqrt{\Omega}.\label{eq:tvp}
\end{equation}
The BCC projector is
\begin{equation}\tilde{X}^\infty = \Omega^\infty Bc\sqrt{\Omega}.\end{equation}
To check the weak \integrability condition, compute
\begin{equation}Q_\Psi \tilde{U}=\sqrt{\Omega}\Big(c-[\Omega,c]\Big)
\frac{KB}{1-\Omega}c\sqrt{\Omega}.\end{equation}
Then
\begin{eqnarray}\tilde{X}^\infty Q_\Psi \tilde{U} \lineup = 
\Omega^\infty Bc\Omega \Big(c-[\Omega,c]\Big)\frac{KB}{1-\Omega}c
\sqrt{\Omega}\nonumber\\
\lineup = \Omega^\infty(1-\Omega)[\Omega,c]\frac{KB}{1-\Omega}c\sqrt{\Omega}
\nonumber\\
\lineup = 0,
\end{eqnarray}
consistently. 

If we compose Okawa's $U$ and $\tilde{U}$, we transform from the perturbative 
vacuum, to the tachyon vacuum, and back. The result is a singular gauge 
transformation from the perturbative vacuum to itself:
\begin{equation}U_0 = U\tilde{U} = 1-\Omega.\label{eq:p.v.}\end{equation}
The BCC projector of $U_0$ is the sliver state:
\begin{equation}X_0^\infty = \Omega^\infty.\label{eq:slv}\end{equation}
This also happens to be the characteristic projector computed in  \cite{Ian}. 
The sliver state can be seen as a surface of string connecting the perturbative
vacuum to itself. Unlike the BCC projector for Okawa's $U$, this does not 
vanish in the Fock space. However, let us 
explain a point of possible confusion: There are many left gauge 
transformations from the perturbative vacuum to itself whose BCC projector 
vanishes identically. The simplest example is $U=1$. What makes the BCC 
projector \eq{slv} nontrivial is that $U_0$ is an {\it exact} left gauge 
transformation,
\begin{equation}U_0=Q\left( B\frac{1-\Omega}{K}\right),
\end{equation}
and, as argued in section \ref{sec:category}, exact left gauge transformations
naturally generate a surface of string connecting the source and target BCFTs 
(which, in this case, happen to be the same). Proper gauge transformations, 
like $U=1$, should not really be viewed as open strings connecting 
solutions---indeed, a proper gauge transformation around one solution is also 
a proper gauge transformation around any other. Accordingly, they do 
not generate physically interesting BCC projectors.

\subsection{Multibranes and Ghost Branes}
\label{subsec:multiple}

Let see what the formalism has to say about the multiple brane and ghost 
brane solutions discussed in  \cite{multiple,multiple2}. These solutions are 
known to suffer from singularities related to their definition as formal gauge 
transformations of the perturbative vacuum \cite{multiple2,multi_bdry,Hata}. 

\begin{figure}
\begin{center}
\resizebox{4.5in}{1.3in}{\includegraphics{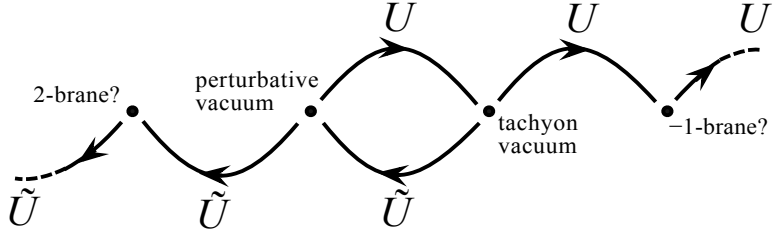}}
\end{center}
\caption{\label{fig:multi} Multiple brane and ghost brane solutions can be
formally defined by iterating the left gauge transformation $U$ from the 
perturbative vacuum to the tachyon vacuum, and the left gauge transformation
$\tilde{U}$ from the tachyon vacuum to the perturbative vacuum.}
\end{figure}

The two-brane solution can be derived by applying Okawa's $\tilde{U}$ in 
\eq{tvp}---which takes the tachyon vacuum to the perturbative vacuum---once 
again to the perturbative vacuum \cite{Ellwood_Schnabl} 
(see figure \ref{fig:multi}). The projector 
$\tilde{X}^\infty = \Omega^\infty Bc\sqrt{\Omega}$ is the same as 
before, but since we are starting from the perturbative vacuum, the weak 
\integrability condition is different:
\begin{eqnarray}\tilde{X}^\infty Q\tilde{U}\lineup
 = -\Omega^\infty Bc\Omega cKBc
\sqrt{\Omega}\nonumber\\
\lineup =-\Omega^\infty c(1-\Omega)KBc\sqrt{\Omega}\nonumber\\
\lineup = B\Omega^\infty cK(1-\Omega)c\sqrt{\Omega}.
\end{eqnarray}
This is not zero. This means that $\tilde{U}$ is not a consistent left gauge 
transformation applied to the perturbative vacuum, and there should be no 
corresponding 2-brane solution. Still the 2-brane solution can be formally 
defined, and recent studies have shown that it is possible to 
recover the correct tension from the action \cite{multiple,multiple2,Hata},  
the closed string tadpole \cite{multiple,multiple2,Ellwood}, 
and, in a limiting case, the boundary state \cite{multi_bdry,boundary}. 
This suggests that there is something essentially ``correct'' about the 
solution which remains to be understood.

Ghost brane solutions are defined by applying Okawa's $U$ in \eq{Okawa} more 
than once to the perturbative vacuum \cite{Ellwood_Schnabl}. They correspond 
to ``removing'' D-branes from an already empty vacuum (for a possible 
interpretation see  \cite{Okuda}). For example, the 
$(-1)$-brane solution can be obtained by applying
\begin{equation}U^2 = 1-\sqrt{\Omega}cB(2-\Omega)\sqrt{
\Omega}\end{equation}
to the perturbative vacuum. We find the BCC projector is
\begin{eqnarray}X^\infty \lineup = 
\sqrt{\Omega}cB\left[\lim_{N\to\infty}\Omega^{N-1}(2-\Omega)^N\right]
\sqrt{\Omega}\nonumber\\
\lineup = \sqrt{\Omega}cB\Omega^\infty,
\end{eqnarray}
which turns out to be the same BCC projector as for $U$ alone.\footnote{It is 
worth noting that the subleading corrections to the sliver state in this limit
are very different from those of a wedge state $\Omega^N$ for large $N$. 
In some contexts the subleading behavior can be physically 
important \cite{exotic}.} To see whether $U^2$ is a sensible left gauge 
transformation, compute
\begin{eqnarray}X^\infty (QU^2) \lineup 
= (X^\infty QU)U+(X^\infty U)QU\nonumber\\
\lineup = 0.\label{eq:weakU2}\end{eqnarray}
Surprisingly, the weak consistency condition does not reveal an obstruction for
negative tension branes! Still, it turns out that $U^2$ is inconsistent. 
To see what's going on, 
consider the equation
\begin{equation}1-\Omega = K^2 M.\label{eq:M2}\end{equation}
which we want to solve for $M$. Multiplying both sides by the sliver state 
consistently gives $0=0$. But still \eq{M2} has no solution because the 
image of $K^2$ is smaller than the image of $1-\Omega$ (which vanishes only 
linearly at $K=0$). In a similar way, while $U^2$ satisfies the weak 
\integrability condition, the BRST charge does not map the image of $U^2$ into
itself.

Actually, there is nothing special going on here with the ghost brane 
solutions. It is a general expectation that the BCC projector for any 
left gauge transformation $U$ should be the same as for $U^N$ for any positive
power $N$. Then it follows that the weak \integrability condition is 
obeyed for any power $U^N$ if it is satisfied for $U$. For example, if we 
had applied Okawa's $\tilde{U}$ twice to the tachyon 
vacuum, instead of once to the perturbative vacuum, the weak \integrability 
condition would also work for the 2-brane solution. Therefore, to 
really test the ghost brane we should apply Okawa's $U$ once to 
the tachyon vacuum, rather than twice to the perturbative vacuum. Then 
the weak \integrability condition takes a different form: 
\begin{eqnarray}X^\infty Q_\Psi U \lineup = 
X^\infty QU + X^\infty [\Psi,U]\nonumber\\
\lineup = 
\sqrt{\Omega}cB\Omega^\infty \left(-c\frac{KB}{1-\Omega}c\Omega cB\sqrt{\Omega}
+cB\Omega c\frac{KB}{1-\Omega}c\sqrt{\Omega}\right)\nonumber\\
\lineup = \sqrt{\Omega}cB\Omega^\infty \left(-c\Omega cB\sqrt{\Omega}
+c\sqrt{\Omega}\right)\nonumber\\
\lineup =\sqrt{\Omega}cB\Omega^\infty c(2-\Omega)\sqrt{\Omega}.\end{eqnarray}
This time we do not find zero. 

It is interesting to speculate what the characteristic projector
might look like for a 2-brane solution. Following the proposal of 
Ellwood \cite{Ian}, one expects that the characteristic projector of  
a solution $\Phi$ describes the boundary conditions of $\Phi$ towards the 
midpoint of the projector. We can generalize this proposal as follows. 
Let $U$ be a left gauge transformation from $\Phi_1$ to 
$\Phi_2$, and assume that $\Phi_1$ is real.\footnote{In general $U$ will not
generate a real solution from $\Phi_1$, so we do not assume $\Phi_2$ is real.}
Then the conjugate gauge parameter $U^\ddag$ is a left gauge transformation 
from $\Phi_2^\ddag$ to $\Phi_1$ and 
\begin{equation}U^\ddag U\end{equation}
is a singular gauge transformation from $\Phi_2^\ddag$ to $\Phi_2$.
Assuming $\Phi_2^\ddag$ and $\Phi_2$ are gauge equivalent, 
the BCC projector for $U^\ddag U$ should describe the boundary 
conditions of the desired BCFT towards the midpoint. For multiple brane 
solutions, these boundary conditions must include Chan-Paton factors, and it 
has been speculated that the Chan-Paton structure should be described by 
higher rank star algebra projectors 
\cite{Ian,RSZsplit,RSZproj,Mac}. On the other hand, applying this argument to 
the formal 2-brane solution constructed from Okawa's $\tilde{U}$ gives 
\begin{equation}\tilde{U}^\ddag\tilde{U}=1-\Omega\end{equation}
whose BCC projector is the sliver state. The sliver state is believed to be 
(in some sense) a rank one projector, so we would not expect this ansatz 
to reproduce the non-abelian Chan-Paton structure of the 2-brane.

\subsection{Ellwood/BMT Lumps}

Following  \cite{Ian}, there has been interest in using singular gauge
transformations to construct solutions describing the endpoint of an RG flow
triggered by a relevant deformation \cite{BMT,lumps,BGT}, which in particular
can describe the tachyon lump \cite{Moeller}. A simple example 
of such a solution was discovered by Bonora, Tolla and one of the authors 
(BMT) and takes the form \cite{BMT}
\begin{equation}\Phi=c\phi - \frac{B}{K+\phi}\phi'c\d c,\label{eq:BMT}
\end{equation}
where $\phi$ is a relevant matter operator in the reference boundary conformal
field theory and $\phi'$ describes the failure of $\phi$ 
to be marginal. The operator $\phi$ must be appropriately ``tuned'' to trigger
an RG flow to the desired boundary conformal field theory ($\mathrm{BCFT}^*$)
in the infrared \cite{Ian,BMT,lumps}.

The solution can be derived from the tachyon vacuum \cite{simple} 
\begin{equation}
\Psi=\frac{1}{1+K}(c+Q(Bc)),
\end{equation}
using a (naive) singular gauge transformation \cite{BMT}
\begin{equation}
U=1-\frac 1{1+K}(1-\phi)Bc.\label{eq:lumpU}
\end{equation}
To derive the BCC projector, we use the formula \eq{Ueps} (with a redefinition
$\eps\to\eps/\beps$):
\begin{equation}X^\infty = \lim_{\eps\to0^+}\frac{\eps}{\eps+\beps U}
,\ \ \ \ \ \ \ \beps\equiv 1-\eps.\label{eq:eps_reg}\end{equation}
This gives
\begin{eqnarray}X^\infty\lineup = \lim_{\eps\to0^+}\left[ 
\frac{\eps}{1-\frac{\beps}{1+K}(1-\phi)Bc}\right]\nonumber\\
\lineup = \lim_{\eps\to0^+}\left[ \eps+
\frac{\eps}{1-\frac{\beps}{1+K}(1-\phi)}
\frac{\beps}{1+K}(1-\phi)Bc\right]\nonumber\\
\lineup = \lim_{\eps\to0^+}\left[
\frac{\eps}{\eps+K+\beps\phi}\right](1-\phi)Bc.
\end{eqnarray}
As shown in  \cite{lumps}, the limit in parentheses converges to the 
so-called {\it deformed sliver state} $\tilde{\Omega}^\infty$, which is 
the sliver state with an insertion of the relevant boundary interaction 
at constant RG coupling in the upper half plane representation. 
Then the BCC projector is
\begin{equation}X^\infty = \tilde{\Omega}^\infty (1-\phi)Bc.\label{eq:lumpX}
\end{equation}
This state vanishes in the Fock space, which is apparently a consequence of 
the fact that the source solution is the tachyon vacuum.

If we map the deformed sliver state to the unit disk 
representation \cite{RSZproj}, the midpoint of the local coordinate touches 
the boundary of the disk, splitting the surface in two. 
Since the conformal transformation is singular at 
the midpoint, the RG coupling of the relevant boundary 
interaction is pushed to the strict infrared, so that the disk correlator has 
$\mathrm{BCFT}^*$ boundary conditions precisely where the midpoint
of the local coordinate touches the boundary of the unit disk. In this sense, 
the BCC projector \eq{lumpX} has $\mathrm{BCFT}^*$ boundary 
conditions at the midpoint. In fact, the naive left gauge transformation 
\eq{lumpU} was constructed 
to give precisely this result, based on the conjecture of Ellwood \cite{Ian}
that the characteristic projector of a solution $\Phi$ should have 
boundary conditions corresponding to the BCFT of $\Phi$ at the midpoint. 
While this idea seems correct, it is not sufficient; 
The BRST variation of $U$ must be proportional to $U$.

To see if this is the case, let's look at the weak \integrability condition. 
We assume \eq{Bneq0} and
the analogue of \eq{Keq0} for the deformed sliver state:
\begin{equation}(K+\phi)\tilde{\Omega}^\infty =0.\label{eq:dKeq0}\end{equation}
Computing
\begin{equation}Q_\Psi U = -(1-\phi)c+\frac{1}{1+K}(K+\phi)\frac{1}{1+K}Bc\d c
-\frac{B}{1+K}\phi'c\d c,\end{equation}
we find
\begin{equation}X^\infty Q_\Psi U = 
-\tilde{\Omega}^\infty(1-\phi)\frac{1}{1+K}\left[(K+\phi)
\frac{1}{1+K}Bc\d c-B\phi'c\d c\right].\end{equation}
Now use \eq{dKeq0} to replace $1-\phi$ with $1+K$. Then there is a 
cancellation with $\frac{1}{K+1}$ and the first term in brackets disappears 
when $K+\phi$ hits the deformed sliver. This leaves
\begin{equation} X^\infty Q_\Psi U= -\tilde{\Omega}^\infty B\phi'c\d c.
\end{equation}
The operator $\phi'$ does not generally annihilate the deformed 
sliver \cite{lumps}, so the weak \integrability condition is violated. 
The only way to avoid this problem is to set $\phi'=0$, in which case 
the BMT solution describes a marginal deformation, or to set 
$\tilde{\Omega}^\infty=0$ in which case the BMT solution describes the 
tachyon vacuum. Otherwise, the BMT solution is singular and does not satisfy
the equations of motion \cite{lumps}. Nevertheless, the solution is 
(in a sense) very close to solving the equations of motion, and if one 
carefully treats its singularities, one can recover almost all of the 
expected physics of the RG flow \cite{lumps,BGT}.

\subsection{Solutions of Kiermaier, Okawa, and Soler}
\label{subsec:KOS}

To see a shift in boundary condition inside the BCC projector, we 
should consider a singular gauge transformation relating two solutions which 
describe distinct backgrounds which support open string states. For this 
purpose, it is useful to study the solutions discovered by 
Kiermaier, Okawa, and Soler \cite{bcc} (the KOS solutions): 
\begin{equation}\Phi = -(c\d\sigma_{01})\frac{1}{1+K}
\sigma_{10}(1+K)Bc\frac{1}{1+K}.\label{eq:KOS}\end{equation}
Here $\sigma_{01}$ and $\sigma_{10}$ are dimension $0$ 
primaries with the property that $\sigma_{01}\sigma_{10}=1$.\footnote{We 
assume that $\sigma_{01}$ and $\sigma_{10}$ are dimension $0$ primaries for
 simplicity. We put the ``security strip'' $\frac{1}{1+K}$ on the right to 
make shorter formulas.} In interesting examples, $\sigma_{01}$ and 
$\sigma_{10}$ are boundary condition changing operators from a reference 
$\mathrm{BCFT}_0$ to the open string background $\mathrm{BCFT}_1$ described 
by the solution. Under these assumptions the KOS solution is 
known to describe nonsingular marginal deformations \cite{bcc,bcc_super}.

We can build the KOS solution with a 
left gauge transformation out of the tachyon vacuum:
\begin{eqnarray}U \lineup = Q_{\Psi\Phi}\left(\frac{B}{1+K}\right)
\nonumber\\
\lineup = 1-\sigma_{01}\frac{1}{1+K}\sigma_{10}(1+K)Bc\frac{1}{1+K},
\label{eq:U1}\end{eqnarray}
where we choose $\Psi$ to be the ``simple'' tachyon vacuum solution 
\cite{simple}:
\begin{equation}\Psi=(c+Q(Bc))\frac{1}{1+K}.\label{eq:simp_tv}\end{equation}
Computing the BCC projector we find 
\begin{equation}X^\infty = \sigma_{01}\Omega^\infty \sigma_{10}(1+K)Bc
\frac{1}{1+K}.\label{eq:X1}\end{equation} 
As expected, this vanishes in the Fock space since there are no open strings 
connecting the tachyon vacuum to the marginally deformed D-brane. 
To check the weak \integrability condition, compute 
\begin{equation}Q_{\Psi}U = 
\left(c (1+K)\sigma_{01}\frac{1}{1+K}\sigma_{10}
+\sigma_{01}\left[c,\frac{1}{1+K}\right]\sigma_{10}\right)(1+K)Bc\frac{1}{1+K},
\label{eq:KOSQ}\end{equation}
and then multiply by the BCC projector:
\begin{eqnarray}X^\infty Q_\Psi U\lineup 
=\left(\sigma_{01}\Omega^\infty B\d c\frac{1}{1+K}\sigma_{10}
-\sigma_{01}\Omega^\infty B\left[c,\frac{1}{1+K}\right]\sigma_{10}
\right)(1+K)c\frac{1}{1+K}
\nonumber\\
\lineup = 
\left(\sigma_{01}\Omega^\infty B c\frac{1}{1+K}\sigma_{10}
-\sigma_{01}\Omega^\infty B c\sigma_{10}\right.\nonumber\\
\lineup \ \ \ \ \ \ \left. 
-\sigma_{01}\Omega^\infty Bc\frac{1}{1+K}\sigma_{10}+
\sigma_{01}\Omega^\infty Bc\sigma_{10}\right)(1+K)c\frac{1}{1+K}\nonumber\\
\lineup =0,\label{eq:KOSweak}
\end{eqnarray}
where in the second step we wrote $\d c = [1+K,c]$ and canceled $K$ against
the sliver. If we had only guessed the left gauge transformation \eq{U1} 
without knowing about the KOS solution from the beginning, the final 
cancellation in \eq{KOSweak} would seem quite miraculous. 
For example, suppose we tried to build a marginal solution like KOS from the 
tachyon vacuum using
\begin{equation}U\? 1-\sigma_{01}\frac{1}{1+K}\sigma_{10}Bc.\end{equation}
At first sight, this guess seems physically plausible. It is non invertible,
and the BCC projector vanishes in the Fock space and has $\mathrm{BCFT}_1$
boundary conditions at the midpoint. Nevertheless,  this
ansatz fails to satisfy the weak \integrability condition. This is an 
important point: Constructing new solutions by a pure gauge ansatz requires 
more than a plausible guess. The structure of the ansatz has to work in a 
nontrivial fashion.

Our main interest in this section is computing a boundary condition changing
projector which displays a shift in boundary condition between two nontrivial 
backgrounds. Accordingly, consider two KOS solutions:
\begin{eqnarray}\Phi_1 \lineup= -(c\d\sigma_{01})\frac{1}{1+K}
\sigma_{10}(1+K)Bc\frac{1}{1+K},\ \ \ \ \ \ \ \ (\mathrm{BCFT}_1),\nonumber\\ 
\Phi_2\lineup = -(c\d\sigma_{02})\frac{1}{1+K}
\sigma_{20}(1+K)Bc\frac{1}{1+K},\ \ \ \ \ \ \ \ (\mathrm{BCFT}_2).
\label{eq:KOS2}\end{eqnarray}
related by the left gauge transformation 
\begin{eqnarray}U_{12} \lineup = Q_{\Phi_1\Phi_2}\left(\frac{B}{1+K}\right)
\nonumber\\ 
\lineup = 1-cB(1+K)\sigma_{01}\frac{1}{1+K}\sigma_{10}\frac{1}{1+K}
-\sigma_{02}\frac{1}{1+K}\sigma_{20}(1+K)Bc\frac{1}{1+K}.
\label{eq:U12}\end{eqnarray}
Note that $U_{12}$ can be factorized into a product of left gauge
transformations passing through the tachyon vacuum \eq{simp_tv}. The reason 
is the following: If $A$ 
is a homotopy operator of a tachyon vacuum solution $\Psi$, and $A^2=0$, 
we can factorize any exact left gauge transformation derived from $A$ into 
a product of two left gauge transformations passing through $\Psi$:
\begin{equation}Q_{\Phi_1\Phi_2}A = (Q_{\Phi_1\Psi}A)(Q_{\Psi\Phi_2}A).
\end{equation}
We do not believe this property is essential for the interpretation of 
the BCC projector, but it would be worth understanding this issue.

To compute the BCC projector we use \eq{eps_reg}
\begin{equation}X_{12}^\infty = \lim_{\eps\to 0^+}
\frac{\eps}{\eps+\beps U_{12}}.\end{equation}
Plugging in $U_{12}$,
\begin{eqnarray}\frac{\eps}{\eps+\beps U_{12}} \lineup = 
\frac{\eps}{1-\beps\left[cB(1+K)\sigma_{01}\frac{1}{1+K}\sigma_{10}
+\sigma_{02}\frac{1}{1+K}\sigma_{20}(1+K)Bc\right]\frac{1}{1+K}}
\nonumber\\ 
\lineup = 
\frac{\eps}{\left(1-\beps cB(1+K)\sigma_{01}\frac{1}{1+K}\sigma_{10}
\frac{1}{K+1}\right)\left(1-\beps\sigma_{02}\frac{1}{1+K}\sigma_{20}(1+K)Bc
\frac{1}{1+K}\right)}\nonumber\\
\lineup = \eps\left(\frac{1}{1-\beps\sigma_{02}\frac{1}{1+K}\sigma_{20}
(1+K)Bc\frac{1}{1+K}}\right)\left(\frac{1}{1
-\beps cB(1+K)\sigma_{01}\frac{1}{1+K}\sigma_{10}
\frac{1}{K+1}}\right)\nonumber\\
\lineup = \eps\left(1+\sigma_{02}\frac{\beps}{K+\eps}\sigma_{20}(1+K)
Bc\frac{1}{1+K}\right)\left(1+cB(1+K)\sigma_{01}
\frac{\beps}{\eps+K}\sigma_{10}\frac{1}{1+K}\right).\nonumber\\
\end{eqnarray}
At this point we can already see the boundary conditions of $\mathrm{BCFT}_2$
gathering on the left and the boundary conditions of $\mathrm{BCFT}_1$ on the 
right. Multiplying this out we get four terms: 
\begin{eqnarray}
\frac{\eps}{\eps+\beps U_{12}} \lineup = \eps 
+ \beps \left(\sigma_{02}\frac{\eps}{\eps+K}\sigma_{20}(1+K)Bc\frac{1}{1+K}
\right)
+ \beps\left(cB(1+K)\sigma_{01}
\frac{\eps}{\eps+K}\sigma_{10}\frac{1}{1+K}\right)\nonumber\\
\lineup \ \ \ \ \ 
+\beps^2\left( \sigma_{02}\frac{\eps}{\eps+K}\sigma_{21}B\d c 
\frac{1}{\eps+K}\sigma_{10}\frac{1}{1+K}\right),
\label{eq:X121}\end{eqnarray}
where $\sigma_{21}\equiv\sigma_{20}\sigma_{01}$ is the boundary condition 
changing operator between $\mathrm{BCFT}_2$ and $\mathrm{BCFT}_1$. For 
simplicity we are assuming that $\sigma_{20}$ and $\sigma_{01}$ have 
trivial contractions. In principle this assumption should not be 
necessary---the collision between $\sigma_{20}$ and $\sigma_{01}$ could be 
vanishing or divergent, depending on whether the BCC operator $\sigma_{21}$ 
has positive or negative conformal dimension. However, this limitation is 
an artifact of the KOS solution and the chosen 
left gauge transformation \eq{U12}. Other choices naturally regulate the 
collision. But we are seeing a precursor to a deeper issue that we will 
discuss shortly. 

\begin{figure}
\begin{center}
\resizebox{4.5in}{2.1in}{\includegraphics{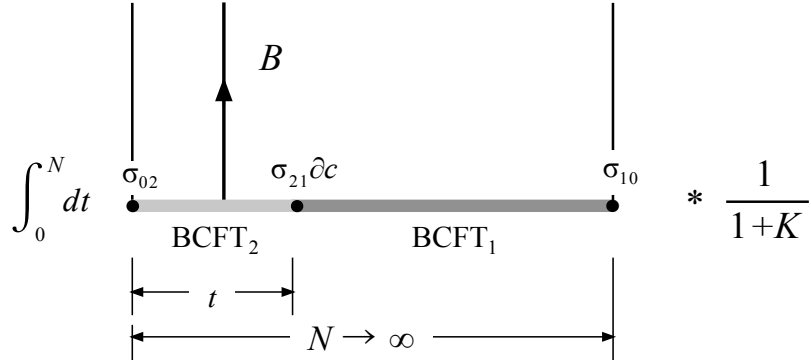}}
\end{center}
\caption{\label{fig:X12} Cylinder representation of the 
nontrivial term in the BCC projector \eq{X12}. The 
integration over $t$ moves the boundary condition changing operator over 
the whole width of the sliver state.}
\end{figure}

To simplify \eq{X121} further, consider the fourth term and 
expand using the Schwinger parameterization:
\begin{equation}\frac{1}{\eps+K}{\sigma_{21}B\d c}\frac{1}{\eps+K}
= \int_0^\infty dt\, e^{-\eps t} \int_0^t ds\, \Omega^{t-s}\sigma_{21}B\d c
\Omega^s.
\end{equation}
Note the identity
\begin{equation}\d\left( \int_0^t ds\, \Omega^{t-s}\Phi \Omega^s\right) 
= [\Phi,\Omega^t].\end{equation}
This suggests that we define the formal expression
\begin{equation}\frac{1}{\d}[\Phi,\Omega^t]\equiv
\int_0^t \Omega^{t-s}\Phi\Omega^s.\end{equation}
Then we can write
\begin{equation}\frac{1}{\eps+K}{\sigma_{21}B\d c}\frac{1}{\eps+K} = 
\frac{1}{\d}\left[\sigma_{21}B\d c,\frac{1}{\eps+K}\right],\end{equation}
and \eq{X121} becomes
\begin{eqnarray}
\frac{\eps}{\eps+\beps U_{12}} \lineup = \eps 
+ \beps\left( \sigma_{02}\frac{\eps}{\eps+K}\sigma_{20}(1+K)Bc\frac{1}{1+K}
\right)
+ \beps\left( cB(1+K)\sigma_{01}
\frac{\eps}{\eps+K}\sigma_{10}\frac{1}{1+K}\right)\nonumber\\
\lineup \ \ \ \ \ 
+ \beps^2\left(\sigma_{02}\frac{1}{\d}\left[\sigma_{21}B\d c,
\frac{\eps}{\eps+K}\right]\sigma_{10}\frac{1}{1+K}\right).\end{eqnarray}
Finally, taking the $\eps\to 0^+$ limit gives:
\begin{eqnarray}\!\!\!\!\!\!\!\!\!\!\!\!\!\!\!\!\!\!\!\!\!\!
X_{12}^\infty
\lineup = \Bigg[\Big(cB(1+K)\,\sigma_{01}\Omega^\infty 
\sigma_{10}\Big)+\Big(\sigma_{02}\Omega^\infty 
\sigma_{20}\,(1+K)Bc \Big)\nonumber\\
\lineup \ \ \ \ \ \ \ \ \ \ \ \ \ \ \ \ \ \ \ \ \ \ \ \ 
\ \ \ \ \ \ \ \ \ \ \ \ \ \ \ \ 
+\Big(\sigma_{02}\,\frac{1}{\d}\big[\sigma_{21}B\d c,
\Omega^\infty \big]\sigma_{10}\Big)\Bigg]\frac{1}{1+K}.
\label{eq:X12}
\end{eqnarray}
The first two terms are BCC projectors for left gauge transformations 
from $\Phi_1$ to the tachyon vacuum \eq{simp_tv}, and from the tachyon vacuum
to $\Phi_2$. These terms vanish in the Fock space. The last term contains 
a shift in boundary condition between $\mathrm{BCFT}_2$ and 
$\mathrm{BCFT}_1$, which is integrated 
(with some ghosts) through the entire width of the sliver state 
(see figure \ref{fig:X12}). This term does not vanish in the 
Fock space.

The BCC projector we just computed should reduce to the characteristic 
projector in the special case when $\mathrm{BCFT}_2=\mathrm{BCFT}_1$. 
Let's see how this happens:
\begin{eqnarray}X_{11}^\infty \lineup 
= \left[cB(1+K)\sigma_{01}\Omega^\infty\sigma_{10}+\sigma_{01}\Omega^\infty \sigma_{10}(1+K)Bc
+\sigma_{01}\frac{1}{\d}\d [Bc,\Omega^\infty ]\sigma_{10}\right]
\frac{1}{1+K}\nonumber\\
\lineup = \Big[cB(1+K)\sigma_{01}\Omega^\infty\sigma_{10}+\sigma_{01}\Omega^\infty \sigma_{10}(1+K)Bc
+Bc\sigma_{01}\Omega^\infty\sigma_{10}-\sigma_{01}\Omega^\infty\sigma_{10}Bc
\Big]\frac{1}{1+K}\nonumber\\
\lineup = \sigma_{01}\Omega^\infty\sigma_{10}\frac{1}{1+K}
+\Big[cBK\sigma_{01}\Omega^\infty\sigma_{10}+\sigma_{01}\Omega^\infty 
\sigma_{10}KBc\Big]\frac{1}{1+K}.
\end{eqnarray}
Since $K$ annihilates the sliver, the $K$s next to the BCC operators in the 
second term can be traded with worldsheet derivatives. Then
\begin{equation} X_{11}^\infty = 
\sigma_{01}\Omega^\infty\sigma_{10}\frac{1}{1+K}
+\Big[(Q\sigma_{01})B\Omega^\infty\sigma_{10}-\sigma_{01}B\Omega^\infty 
(Q\sigma_{10})\Big]\frac{1}{1+K},\end{equation}
and the characteristic projector simplifies to
\begin{equation}X_{11}^\infty = \Bigg[\sigma_{01}\Omega^\infty\sigma_{10}
+Q\Big(\sigma_{01}B\Omega^\infty \sigma_{10}\Big)\Bigg]\frac{1}{1+K}.
\end{equation}
Extrapolating from Schnabl gauge, this expression agrees with the 
characteristic projector computed in  \cite{Ian} except for 
the second term, which would have been ignored in \cite{Ian} because it 
vanishes in the Fock space. Further specifying 
$\mathrm{BCFT}_1=\mathrm{BCFT}_0$ gives $X_{00}^\infty=\Omega^\infty$, which 
is the BCC projector for a singular gauge transformation from the 
perturbative vacuum to itself.

\begin{figure}
\begin{center}
\resizebox{2.4in}{2.4in}{\includegraphics{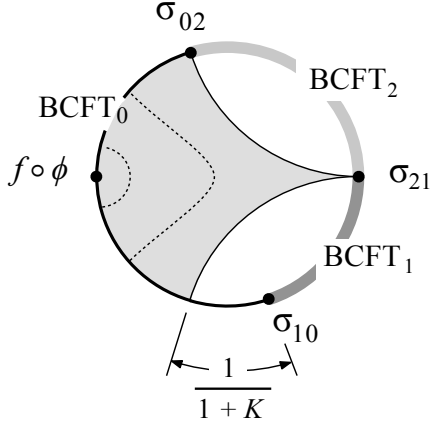}}
\end{center}
\caption{\label{fig:BCCKOS} The BCC projector \eq{X12} contracted with a 
Fock space state, represented as a correlator on the unit disk. Note
that the change of boundary condition happens precisely at the point where 
the midpoint of the local coordinate touches the boundary of the disk. }
\end{figure}

The form of the BCC projector $X_{12}^\infty$ substantially simplifies if 
we only contract with regular test states. Then first two terms in \eq{X12}
can be ignored, and the third term simplifies to: 
\begin{equation}X_{12}^\infty = \Big[\sigma_{02}\Omega^\infty \sigma_{21}
\Omega^\infty \sigma_{10}\Big]\frac{1}{1+K},\ \ \ \ \ \ (\mathrm{Fock\ space})
.\label{eq:Fock_X12}\end{equation}
For the detailed argument, see appendix \ref{app:KOS_BCC}.
This is precisely the structure we expected: $X_{12}^\infty$ is 
proportional to the sliver state with the boundary conditions of 
$\mathrm{BCFT}_2$ on its left half and the boundary conditions of 
$\mathrm{BCFT}_1$ on its right half, with the BCC operator $\sigma_{21}$ 
inserted at the midpoint (see figure \ref{fig:BCCKOS}).

The KOS solutions are a special case of the solutions for regular marginal 
deformations in dressed Schnabl-gauges \cite{simple,supermarg}. Let us see 
how this discussion extends to the Schnabl-gauge marginal 
solutions \cite{RZOK,Schnabl2}:
\begin{eqnarray}\Phi_1 \lineup 
= \sqrt{\Omega}\, cV_1\frac{1}{1+\frac{1-\Omega}{K}V_1}Bc\, \sqrt{\Omega},
\ \ \ \ \ \ \ \ \ \ \ (\mathrm{BCFT}_1),\nonumber\\
\Phi_2 \lineup 
= \sqrt{\Omega}\, cV_2\frac{1}{1+\frac{1-\Omega}{K}V_2}Bc\,\sqrt{\Omega},
\ \ \ \ \ \ \ \ \ \ \ (\mathrm{BCFT}_2),
\end{eqnarray} 
where $V_1,V_2$ are nonsingular marginal currents. These solutions can 
be related by a left gauge transformation
\begin{eqnarray}U_{12} \lineup 
= Q_{\Phi_1\Phi_2}\left(B\frac{1-\Omega}{K}\right)
\nonumber\\
\lineup = 1-\sqrt{\Omega}\,cB \frac{1}{1+V_1\frac{1-\Omega}{K}}\, \sqrt{\Omega}
 - \sqrt{\Omega}\frac{1}{1+\frac{1-\Omega}{K}V_2}Bc\, \sqrt{\Omega}.
\end{eqnarray}
This left gauge transformation factorizes through Schnabl's solution for 
the tachyon vacuum, \eq{Schnabl}. 
Computing the BCC projector, we find three terms analogous to \eq{X12}:
\begin{eqnarray}X_{12}^\infty \lineup\ =\ 
\sqrt{\Omega}\Bigg[\left(cB\frac{K}{1-\Omega}
\sigma_{01}\Omega^\infty
\sigma_{10}\right) +\left(\sigma_{02}\Omega^\infty \sigma_{20}
\frac{K}{1-\Omega}Bc\right) 
\nonumber\\ \lineup\ \ \ \ \ \ \ \ \ \ \ \ \ \ \ \ \ \ \ \ \  
+ \left(\sigma_{02}\frac{1}{\d}\left[
\sigma_{20}\frac{K}{1-\Omega}Bc\Omega cB\frac{K}{1-\Omega}\sigma_{01}
,\Omega^\infty\right]\sigma_{10}\right)\Bigg] \sqrt{\Omega}.
\end{eqnarray}
The first two terms are BCC projectors for singular gauge transformations
from $\Phi_1$ to the tachyon vacuum and from the tachyon vacuum to $\Phi_2$.
In the third term, note that $\sigma_{20}$ and $\sigma_{01}$ are 
separated by a finite region of undeformed surface,
\begin{equation}\sigma_{20}\frac{K}{1-\Omega}Bc\Omega cB\frac{K}{1-\Omega}
\sigma_{01},\end{equation}
unlike with KOS, where they collide to form $\sigma_{21}$. 
However, in the projector limit the collision between $\sigma_{10}$ and 
$\sigma_{02}$ effectively re-emerges since the surface separating them is 
squeezed to vanishing width at the midpoint:
\begin{equation}X_{12}^\infty 
= \sqrt{\Omega}\,\sigma_{02}\Omega^\infty\sigma_{21}
\Omega^\infty\sigma_{10}\,\sqrt{\Omega}, 
\ \ \ \ \ \ (\mathrm{Fock\ space}).\end{equation}
Now if $\sigma_{21}$ has nonzero conformal dimension, the BCC projector will
be vanishing or infinite in the Fock space as a result of the singular 
conformal transformation of $\sigma_{21}$. This problem 
is most likely generic. Between any pair of BCFTs, the boundary condition 
changing operator generally carries nonzero conformal weight, and if the BCC 
projector places this operator at the midpoint, there will be a vanishing or
divergent factor when contracting with Fock space states. 

This problem should be irrelevant if we are using the BCC projector inside 
the phantom term to compute physical observables \cite{phantom}. Still we 
would like to see the BCC projector shift the boundary condition in a 
nonsingular fashion when contracted with regular test states. One way around 
this problem is to put the operator $\sigma_{21}$ on-shell, that is, replace 
$\sigma_{21}\to c\mathcal{V}_{21}$, where $c\mathcal{V}_{21}$ is a dimension 
zero element of the cohomology of $Q$ for a stretched string between the 
BCFTs of $\Phi_2$ and $\Phi_1$. Then the 
``on-shell'' BCC projector takes the form
\begin{equation}X_{\mathcal{V}_{21}}^\infty = \sqrt{\Omega}\,\sigma_{20}
\Omega^\infty\,c\mathcal{V}_{21}\,\Omega^\infty \sigma_{10}\,\sqrt{\Omega}.
\end{equation}
This is now a nonsingular state in the Fock space, and is a projector-like 
representative of the ghost number 1 cohomology of $Q_{\Phi_1\Phi_2}$.
In a much more physical sense than the bare BCC projector, 
$X^\infty_{\mathcal{V}_{12}}$ represents a stretched string connecting two 
BCFTs. Note that BCC projectors around the tachyon vacuum cannot be 
made nonvanishing in this way, since there are no on-shell states to insert
at the midpoint. One shortcoming of this picture is that we would like 
to define the on-shell BCC projector as a limit of a sequence of regular 
states, rather than by hand after the projector limit has been taken. We 
will leave this problem to future work.

\section{Summary and Discussion}

The results of this paper can be summarized by three central ideas:
\begin{description}
\item{\bf 1)} Any two classical solutions in open string field theory can 
be related by a {\it left gauge transformation}, i.e. formal gauge 
transformation defined by a finite gauge parameter $U$ possibly without an 
inverse. Left gauge transformations define the morphisms of a category 
{\bf Left}, whose objects are classical solutions.
\item{\bf 2)} Given any left gauge transformation connecting solutions 
$\Phi_1$ and $\Phi_2$, one can define a star algebra projector which 
describes a stretched string connecting the BCFTs of $\Phi_1$ and $\Phi_2$. 
We call this the {\it boundary condition changing projector}. 
The boundary condition changing projector allows us to naturally associate 
the morphisms of {\bf Left} with open strings connecting D-branes.
It also determines the nontrivial part of the mysterious 
{\it phantom term}, as 
will be described in subsequent work \cite{phantom}.
\item{\bf 3)} Given any left gauge transformation $U$ from a 
source solution $\Phi_1$ to any target solution $\Phi_2$, 
the kinetic operator around $\Phi_1$ maps the image of $U$ into itself. 
This observation
can be used to constrain the possible $U$s which define consistent left gauge 
transformations, as summarized by the 
{\it strong} and {\it weak} \integrability conditions, \eq{int1a} and 
\eq{int2aa}. If one wants to construct a new solution using singular gauge 
transformations, one must be sure that the proposed $U$ satisfies these 
consistency conditions. 
\end{description}
We have tried to provide a clear account of these ideas and their motivation,
though there are many aspects of this picture which remain unexplored and 
we hope can be clarified in future work. Our understanding of the BCC 
projector, in particular, is preliminary. It would be desirable to extend 
our analysis to more examples, especially for solutions describing singular 
marginal deformations \cite{FK,KO,FKsuper,KOsuper}. Some aspects of these 
ideas could in principle be tested in the level expansion. For example, one 
could construct (numerically) a singular 
gauge transformation from the perturbative vacuum to the Siegel gauge tachyon 
condensate \cite{SZ,Taylor,Exp,Jap}, and verify that its BCC projector tends
to vanish in the level expansion, but still nontrivially computes 
the shift in the closed string tadpole amplitude between the perturbative 
vacuum and the tachyon vacuum (via the Siegel gauge analogue of the  
phantom term for Schnabl's solution).

A central assumption of our work is that the image and kernel of a 
left gauge transformation are linearly independent and span the whole space, 
and therefore define a boundary condition changing 
projector. However, it is not difficult to find ghost number zero string
fields where this assumption is false, for example
\begin{equation}U = B\d c\label{eq:Uq}\end{equation}
This is nilpotent, so its kernel and image are not linearly 
independent. However, we are not aware of any genuine left gauge 
transformations of this type. That does not mean they do not exist, and 
do not have an important role to play in the construction of classical 
solutions. A particularly 
interesting possibility is that solutions could be constructed with left 
gauge transformations that are non-unitary isometries 
\cite{partial,partialSFT}.

It would be interesting to see if our results have some generalization to 
other string field theories, especially nonpolynomial superstring field 
theory \cite{Berkovits,WZWBV} and open string field theories based on homotopy 
associative algebras \cite{Zwiebach}. It is even possible that these ideas 
have some realization in closed string field theory \cite{closed}, but the 
analogy between left gauge transformations and open strings suggests that a 
different type of structure might be needed there. It would also be interesting
to see if left gauge transformations can be related to the derived 
category of the topological B-model \cite{Douglas,Sharpe,AL}. If so, 
the category {\bf Left} could provide a setting for understanding D-brane 
charges in string field theory.

The main motivation for our work is the idea that singular gauge 
transformations could eventually provide a systematic construction of
analytic solutions in open string field theory corresponding to any choice 
of BCFT. Much more work remains, but we hope the present contribution is a 
useful step in this direction.

\bigskip

\noindent {\bf Acknowledgments}

\bigskip

\noindent We are grateful to K. Bering, I. Ellwood, Y. Okawa, and B. Wecht for 
useful discussions, and M. Schnabl for discussion and helpful comments on the 
manuscript. T.E. gratefully acknowledges travel support from a joint 
JSPS-MSMT grant LH11106. C.M.  acknowledges INFN sezione di Torino, gruppo 
collegato di Alessandria, for kind partial support. This research was 
funded by the EURYI grant GACR EYI/07/E010 from EUROHORC and ESF.

\begin{appendix}
\section{KOS Projector in Fock space}
\label{app:KOS_BCC}

\begin{figure}
\begin{center}
\resizebox{6in}{2in}{\includegraphics{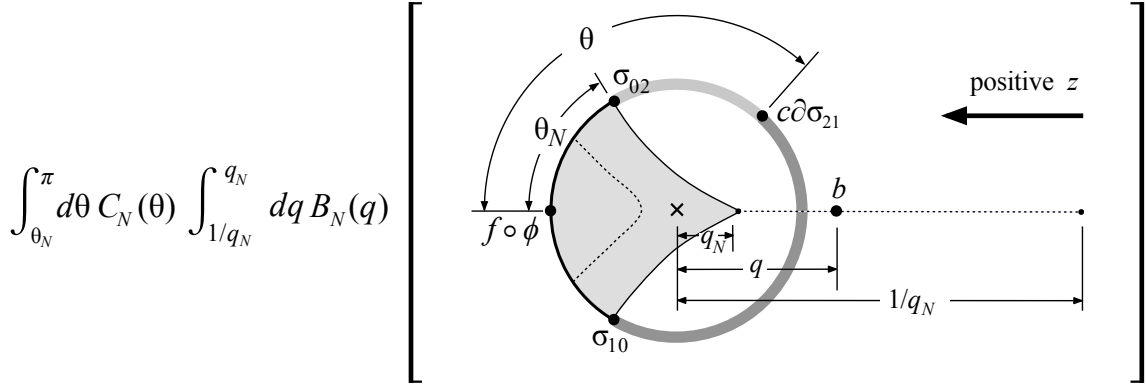}}
\end{center}
\caption{\label{fig:BCCKOS2} Pictorial representation of the correlator 
\eq{van} on the unit disk. The picture is drawn (following the left handed 
convention \cite{simple}) with the real axis increasing to the left.}
\end{figure}

In this appendix we would like to describe how the BCC projector for the KOS 
solution \eq{X12} appears when contracted with Fock space states. The first
two terms in \eq{X12} can be ignored, since they are BCC projectors to and 
from the tachyon vacuum and manifestly vanish in the Fock space. The 
interesting term is the third, which contains the BCC operator $\sigma_{21}$.
Regulating the sliver state, $\Omega^\infty\to\Omega^{2N}$, this term contains 
the factor
\begin{equation}F = \sigma_{02}\frac{1}{\d}\big[B\d c\sigma_{21},
\Omega^{2N}\big]\sigma_{10}.\end{equation}
Split this into a sum of two terms as follows:
\begin{equation}F = -\sigma_{02}\frac{1}{\d}\big[\d c\sigma_{21},
\Omega^{N}\big]B\Omega^N\sigma_{10} + 
\sigma_{02}\Omega^N B \frac{1}{\d}\big[\d c\sigma_{21},
\Omega^{N}\big]\sigma_{10},\label{eq:S1}
\end{equation}
and then write 
\begin{equation}\d c  \sigma_{21} = \d(c\sigma_{21}) - c\d\sigma_{21}.
\end{equation}
Substituting in \eq{S1}, the inverse $\d$ cancels with $\d(c\sigma_{21})$, 
giving
\begin{eqnarray}F \lineup = \sigma_{02}\Omega^N \sigma_{21}
\Omega^N\sigma_{10}- \underbrace{\phantom{\bigg(\!\!\!\!}
\sigma_{01}cB\Omega^{2N}\sigma_{10} 
+\sigma_{02}\Omega^{2N}Bc\sigma_{20}}_{\displaystyle \lim_{N\to\infty}= 0}\nonumber\\
\lineup\ \ + \underbrace{\phantom{\bigg(\!\!\!\!}
\sigma_{02}\frac{1}{\d}\big[ c\d\sigma_{21},
\Omega^{N}\big]B\Omega^N\sigma_{10} - 
\sigma_{02}\Omega^N B \frac{1}{\d}\big[ c\d\sigma_{21},\Omega^{N}\big]
\sigma_{10}}_{\displaystyle \lim_{N\to\infty} =\ ?}.
\label{eq:s1}\end{eqnarray}
The first term gives the claimed Fock space expansion of the BCC projector. 
The second two terms are easily seen to vanish in the Fock space when 
$N\to\infty$. The last two terms (above the question mark) are more 
complicated and it is not obvious what happens to them in the 
$N\to\infty$ limit. On the one hand, $B$ acts on the sliver, which tends to 
make these contributions vanish, but on the other hand there is a divergent 
integration of $c\d \sigma_{21} = Q\sigma_{21}$ on the left or the right half
of the sliver state. To see what happens, we map these terms to the unit 
disk in the representation where the local coordinate patch for the sliver 
state is regular \cite{RSZproj}. In this case, all operator insertions on the 
disk are at finite and nonzero separation (see figure \ref{fig:BCCKOS2}), 
and to analyze the $N\to\infty$ limit all we have to do is carefully keep 
track of the factors which appear from the conformal transformation from the 
cylinder. Then the first term above the question mark in \eq{s1} takes 
the form:
\begin{eqnarray}\lineup\!\!\!\!\!\!\!\!\!\!\!
\left\langle\phi,\sigma_{02}\frac{1}{\d}\big[ c\d\sigma_{21},
\Omega^{N}\big]B\Omega^N\sigma_{10}\right\rangle = \nonumber\\
\lineup \ \ \int_{\theta_N}^\pi 
\frac{d\theta}{2\pi}
C_N(\theta)\int_{1/q_N}^{q_N}dq\,B_N(q)\Bigg\langle
f\circ\phi(0)\ \sigma_{02}(e^{i\theta_N})\,
\Big[\,c\d\sigma_{21}(e^{i\theta})\,b(q)\,
\Big]\,\sigma_{01}(e^{-i\theta_N})\Bigg\rangle_\mathrm{disk}.\nonumber\\
\label{eq:van}\end{eqnarray}

\noindent The quantities appearing in the above correlator are
\begin{eqnarray}C_N(\theta) \lineup 
= \frac{2\sec^2\frac{\theta}{2}}{1+\left(\frac{2}{L}\right)^2
\tan^2\frac{\theta}{2}},\\
B_N(q) \lineup 
= \frac{i}{2L^2}\left[\left(\frac{L}{2}\right)^2(q+1)^2-(q-1)^2
\right],\\
\theta_N \lineup 
= \tan^{-1}\left(\frac{4L\tan\frac{\pi}{2L}}{4-L^2
\tan^2\frac{\pi}{2L}}\right),\\
q_N\lineup = -\frac{L-2}{L+2}.
\end{eqnarray}
where $L\equiv 2N+1$. For large $L$, the integration over $q$ only has support
over a vanishingly small line segment close to $q=-1$, which tends to make 
the state vanish. However, the integration over $\theta$ has a divergence near 
$e^{i\theta}=-1$, where $C_N(\theta)$ has a peak of height 
$\frac{L^2}{2}$, and moreover a further divergence is produced from the 
singular OPE of $b$ and $c$ near $q=e^{i\theta}=-1$. This is the anticipated 
competition between the divergence of the integration of $c\d \sigma_{21}$ 
and the vanishing factor produced when $B$ acts on the sliver. To isolate 
the overall behavior we make a substitution:
\begin{eqnarray}
t \lineup \equiv L(q+1),\nonumber\\
s \lineup \equiv L(\pi-\theta).
\end{eqnarray}
Focusing on the leading large $L$ behavior, \eq{van} becomes
\begin{eqnarray}\lineup \!\!\!\!\!\!\!\!\!\!
\left\langle\phi,\sigma_{02}\frac{1}{\d}\big[ c\d\sigma_{21},
\Omega^{N}\big]B\Omega^N\sigma_{10}\right\rangle = 
\frac{i}{4L^2}\int_0^{L(\pi-\theta_\infty)}\frac{ds}{2\pi}\int_{-4}^4
dt\,\left[\frac{t^2}{4}-4\right]\frac{1}{1+(\frac{s}{4})^2}
\nonumber\\
\lineup\!\!\! 
\times \left\langle f\circ\phi(0)\ \sigma_{02}(e^{i\theta_\infty})\,
\left[\,c\d\sigma_{21}\left(-e^{-is/L}\right)\,
b\left(-1+\frac{t}{L}\right)\,\right]\,\sigma_{01}(e^{-i\theta_\infty})
\right\rangle_\mathrm{disk}+\mathrm{subleading}.\nonumber\\
\end{eqnarray}
The matter component of the correlator is regular when $L\to\infty$. The ghost
component has two terms: one from contractions of $b,c$ with the probe vertex
operator $\phi$, and another from contractions between $b$ and $c$. The former
term is regular for large $L$; since the overall expression is multiplied
by $1/L^2$, it does not contribute in the limit. The latter term however 
has a singularity from the collision of $b$ and $c$ in the $L\to\infty$ limit.
Computing the OPE we find
\begin{eqnarray}\lineup \!\!\!\!\!\!\!\!\!\!\!\!\!
\left\langle\phi,\sigma_{02}\frac{1}{\d}\big[ c\d\sigma_{21},
\Omega^{N}\big]B\Omega^N\sigma_{10}\right\rangle = 
\frac{i}{4L^2}\int_0^{\infty}\frac{ds}{2\pi}\int_{-4}^4
dt\,\left[\frac{t^2}{4}-4\right]\frac{1}{1+(\frac{s}{4})^2}
\nonumber\\
\lineup \times \underbrace{\ \frac{1}{\frac{is}{L}-\frac{t}{L}}\ }_{\mbox{OPE}}
\Big\langle f\circ\phi(0)\ \sigma_{02}(e^{i\theta_\infty})\,
\d\sigma_{21}(-1)\,\sigma_{01}(e^{-i\theta_\infty})
\Big\rangle_\mathrm{disk}+\mathrm{subleading},\nonumber\\
\end{eqnarray}
where the term above the braces comes from the OPE. Computing
the remaining correlator produces an overall constant, which leaves
\begin{eqnarray}
\left\langle\phi,\sigma_{02}\frac{1}{\d}\big[ c\d\sigma_{21},
\Omega^{N}\big]B\Omega^N\sigma_{10}\right\rangle \lineup = 
\frac{\mathrm{(constant)}}{L}\int_0^{\infty}\frac{ds}{2\pi}\int_{-4}^4
dt\,\left[\frac{t^2}{4}-4\right]\frac{1}{1+(\frac{s}{4})^2}
\frac{1}{is-t}\nonumber\\ \lineup \ \ \ \ \ \ \ \ \ \ \ \ \ 
+\mathrm{subleading}.
\end{eqnarray}
The pole from the $bc$ OPE at $s=t$ is integrable, so in total the state 
vanishes as $1/L$. A similar argument applies for the second term above 
the question mark in \eq{s1}. Therefore we recover the claimed Fock space
expansion of the BCC projector.

\end{appendix}

\end{document}